\documentclass[aps,prl,twocolumn,superscriptaddress,nofootinbib]{revtex4-1}
\usepackage{amsmath}
\usepackage{graphicx}
\usepackage{amsbsy}
\usepackage{amssymb}
\usepackage{palatino,avant,graphicx,color}
\usepackage[none]{hyphenat}
\usepackage{tikz}
\usepackage{xy}
\usepackage{comment}
\usepackage[none]{changes}
\usepackage{wrapfig}
\usepackage{float}
\usepackage{mciteplus}
\usepackage{tabularx}
\usepackage{natbib}
\usepackage{hyperref}
\usepackage{mathtools}
\usepackage[artemisia]{textgreek}
\usepackage{newunicodechar}

\newcommand{\cmmnt}[1]{}

\begin{document}

\title{Nonlinear, Tunable and Active Optical Metasurface with Liquid Film}

\author{Shimon Rubin}
\email{rubin.shim@gmail.com}  
\author{Yeshaiahu Fainman}
\affiliation{Department of Electrical and Computer Engineering, University of California, San Diego, 9500 Gilman Dr., La Jolla, California 92023, USA}

\begin{abstract}
Optical metamaterials and metasurfaces which emerged in the course of the last few decades
have revolutionized our understanding of light and light-matter interaction. While solid materials
are naturally employed as key building elements for construction of optical metamaterials mainly
due to their structural stability, practically no attention was given to study of liquid-made optical
2D metasurfaces and the underlying interaction regimes between surface optical modes and
liquids. In this work, we theoretically demonstrate that surface plasmon polaritons and slab
waveguide modes that propagate within a thin liquid dielectric film, trigger optical self-induced
interaction facilitated by surface tension effects, which lead to formation of 2D optical liquid-made
lattices/metasurfaces with tunable symmetry and which can be leveraged for tuning of lasing
modes. Furthermore, we show that the symmetry breaking of the 2D optical liquid lattice leads to
phase transition and tuning of its topological properties which allows to form, destruct and move
Dirac-points in the k-space. Our results indicate that optical liquid lattices support extremely low lasing threshold relative to solid dielectric films and have the potential to serve as configurable analogous computation platform.
\end{abstract}

\maketitle




Metamaterials are composite man-made or natural materials  that possess emergent optical properties which stem from specific spatial arrangements of the constituent sub-wavelength basic units \cite{yablonovitch1987,john1987}, and lead to  
numerous effects such as formation of bandgap in dielectric photonic crystals \cite{bianco2000large,ho1990existence}, suppression of plasmon frequency in metallic meta-materials \cite{pendry1996extremely}, control over the radiation dynamics of embedded active materials \cite{yablonovitch1991donor,painter1999} (see also \cite{notomi2010manipulating} and references within), wave front control using thin elements \cite{yu2014flat}, polarization and phase control in both transmission \cite{arbabi2015dielectric} and reflection \cite{smalley2017luminescent} modes, and control of light properties in low-loss and high-index dielectric resonant Mie nanoparticles \cite{kuznetsov2016optically}. 
Dynamical tuning of optical metamaterials' properties is particularly appealing as it allows to study new regimes and effects of light-matter interaction, 
holds promise for future metamaterial-based devices with novel functionalities achieved by structuring matter on the sub-wavelength scale \cite{shadrivov2015tunable}, and may be also of interest as platforms to simulate many body quantum effects which are challenging to realize in real quantum systems \cite{longhi2009quantum}.
In particular, numerous studies, partly fueled by seminal advances in condensed matter physics such as the discovery of graphene \cite{neto2009ah}, revealed analogies between propagation of light in photonic crystals to dynamics of relativistic Dirac fermions near Dirac points in crystals, 
and dynamics of electrons in topological insulators. 
In the case of 2D photonic structures, Dirac cones have attracted a significant interest because of the existence of robust surface states due to the breaking of parity and time-reversal symmetry \cite{raghu2008} and intrigue transport properties such as  pseudo-diffusive transmittance \cite{sepkhanov2007extremal}, persistence of the Klein effect \cite{bahat2010klein} and breakdown of conical diffraction due to symmetry breaking of the hexagonal symmetry \cite{bahat2008symmetry} or nonlinear interactions \cite{peleg2007conical,ablowitz2009conical}.
Furthermore, 2D periodic structures play an important role in realizing lasing effects in the so-called distributed feedback (DFB) structures, which provide continuous coherent back-scattering from the periodic structures without mirrors; these were originally proposed in 1D \cite{kogelnik1971stimulated,shank1971tunable}, and later were extended to 2D systems enabling lasing amplification of waveguide modes in photonic crystals \cite{meier1999laser} and of surface plasmon polaritons (SPPs) oscillations in metal structures \cite{zhou2013lasing},
and more recently also in thin polymer membranes \cite{karl2018flexible}. 
Since metasurfaces in general and 2D periodic structures in particular, are conventionally constructed using solid metals and high index dielectrics, their tuning properties are constrained by physical properties such as carrier \cite{ju2011graphene} and material density which can be manipulated by external electrical, magnetic, acoustic and temperature fields \cite{xiao2009tunable}, and also by mechanical stretching of elastic DFB structure which affects the resonant lasing frequencies \cite{zhang2006mechanically}.

Liquids on the other hand, provide an attractive platform to introduce significant changes to the optical properties of metamaterials and metasurfaces, due to the liquids' capability to induce relatively large dynamical changes of their dielectric function, that stem from their adaptive property to fill micro-channels of desired shape \cite{psaltis2006developing}, compliance under external stimuli, and the ability to sustain changes of their physico-chemical properties. Prominent examples include light-induced collective orientation of liquid crystal molecules \cite{busch1999liquid}, pressure induced control of lasing frequency in a 1D array of droplets \cite{bakal2015tunable}, magnetic induced ferrofluid-based hyperbolic metamaterial \cite{smolyaninova2014self}, and chemical composition induced changes enabling lasing frequency tuning in 1D systems \cite{gersborg2007tunability} and in 2D photonic crystals \cite{bernal2007optofluidic}. Furthermore, recent studies introduced a novel 
interaction between SPPs and a thin liquid dielectric (TLD) film due to geometrical changes of the gas-fluid (or fluid-fluid) interface, facilitated by the thermocapillary (TC) effect; theoretical study of self-induced focusing and defocusing effects of propagating SPPs due to nonlocal interaction \cite{rubin2018nonlocal} (where refractive index changes extend beyond the regions of maximal optical intensity) as well as formation of an optical liquid lattice of a fixed square symmetry, and experimental demonstration of TC-assisted optical tuning of surface plasmon resonance coupling angle \cite{rubin2019subnanometer}. While these two studies demonstrate a significant coupling between the topography of the gas-fluid interface and propagation of SPPs modes, these did not leverage surface tension nor surface optical modes to study bandstructure and lasing modes tuning 
due to optically induced changes of the liquid lattices symmetry.

 \begin{figure*}
	\includegraphics[width=\textwidth]{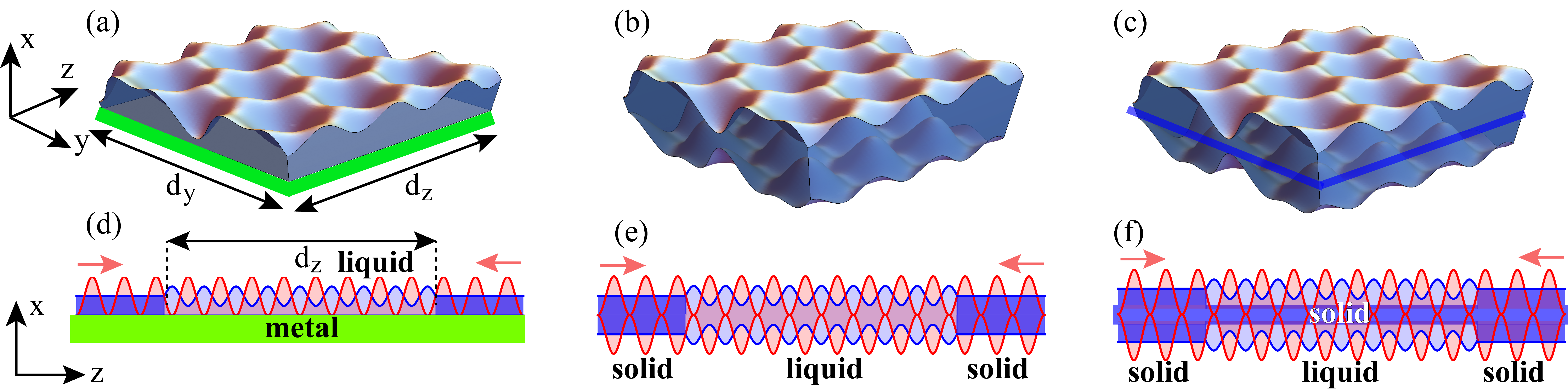}
    \caption{Schematic presentation of TLD film deformation forming optical liquid lattices (blue) due to surface tension effects triggered by interference of surface optical modes (red). (a) 2D Plasmonic liquid lattice formed by interference of SPPs, (b,c) suspended and supported photonic liquid lattice, respectively, formed by interference of photonic slab WG modes. Gain can be introduced either to the liquid or to the solid substrate described in (c). The lateral dimensions of the liquid slots, which are bounded by solid dielectric walls (not shown) are $d_{y}$ and $d_{z}$. (d-f) The corresponding 1D optical liquid lattices in a liquid slot of length $d_{z}$, induced by a pairs of counter-propagating SPPs (d) or slab WG modes (e,f).}
    \label{Setup}
\end{figure*}

In this work, we theoretically demonstrate that SPPs or slab waveguide (WG) modes propagating in a TLD film, which is thinner than the penetration depth
of the corresponding surface optical mode in the direction normal to the film's surface, lead to a nonlocal and nonlinear response of the corresponding dielectric function due to optically-driven surface tension effects. In particular, we take advantage of the surface tension dependence on local physico-chemical conditions \cite{levich1962physicochemical}, to show that the optically induced TC \cite{wedershoven2014infrared} or solutocapillary (SC) effects \cite{muller2017monitoring}, which stem from temperature or chemical concentration gradients, respectively, 
lead to self-induced changes of the dielectric function. In particular, since both surface tension effects are accompanied by flows and deformation of the TLD film, the latter lead to self-induced changes of the dielectric function that are coupled back to the propagation conditions of the surface optical mode.
Importantly, the fact that both SPP and WG modes can propagate within thin dielectric films, which are thinner than the optical penetration depth of the corresponding surface optical mode into the domain outside the TLD film, facilitates a significant coupling between these surface optical modes and changes of liquid's topography. 
In particular, the SPPs and slab WG modes are both guaranteed to propagate in arbitrarily thin liquid films; the former due to the fundamental capability of metal-dielectric interfaces to support SPPs \cite{raether2013surface} whereas the latter can be described by the transverse resonance condition of the fundamental mode \cite{kogelnik1975theory}. 
Employing this interaction for the case of interfering surface optical waves, leads to a self-induced optical liquid lattice of tunable symmetry and bandstructure, which can be tuned by changing the relative propagation directions and the amplitudes of the interfering SPP or WG modes. Furthermore, applying bandstructure tuning in the case the TLD film or the dielectric substrate admit gain properties, lead to configurable DFB mechanism with gain and/or index modulation that can control the threshold condition and the corresponding lasing frequency. In particular, dielectric substrate with gain and TLD film without gain lead to index modulation, whereas the complementary case of TLD film with gain supports both index and gain modulation, as shortly discussed below. 
 
Fig.\ref{Setup} presents a periodic deformation of a TLD film driven by SPPs propagating on a metal-fluid interface and forming a plasmonic liquid lattice, whereas and Fig.\ref{Setup}(b) presents a periodic deformation of a pair of gas-fluid interfaces present in a symmetric liquid slab WG due to propagating WG modes forming a suspended photonic liquid lattice without metals. In practice, the suspended optically thin liquid film can be realized either by bracketing it with an immiscible liquid which admits a distinct refractive index \cite{yaminksi2010stability} or by stabilizing liquid film with surfactants leading to optically thin and stable liquid sheet with a pair of parallel gas-fluid interfaces (see also \cite{patsyk2019observation} for a recent experiments of optical guiding in soap films). Fig.\ref{Setup}(c) presents a thin solid dielectric substrate which hosts a TLD film on both of its sides, leading to a symmetric slab WG forming a supported photonic liquid lattice. 
For simplicity through the work we consider the case where the refractive index of the solid dielectric components and of the TLD film are equal to minimize the impedance mismatch, which could be realized by employing a silicone oil and silica of refractive indices $1.4$ and $1.45$ (under $785$ nm illumination), respectively. 
Under this assumption the system described in Fig.\ref{Setup}(c) allows in principle to realize the optically similar setup presented in Fig.\ref{Setup}(b), without employing surfactants, provided the film deformation is sufficiently small. To the best of our knowledge the prospect of employing the prominent interaction of surface optical modes and TLD film, in order to induce and control the symmetry of 2D optical liquid lattices, and furthermore to leverage it for lasing modes and of topological properties tuning, was not explored to date.

This manuscript is structured as follows. First, we describe the nonlinear self-induced interaction between surface optical modes and the TLD film driven by either the TC or SC effects, and derive the underlying complex nonlocal Ginzburg-Landau equation that governs the dynamics of the corresponding envelope function. 
We then solve the TLD film equation, describe tuning of the optical liquid lattices and of the corresponding symmetries due to changes of the propagation directions and amplitudes of the surface optical modes. 
As an illustrative example we first consider the simpler 1D case and analyze the threshold change and lasing frequency tuning, due to self-induced and non self-induced amplitude modulation of the formed liquid lattice. Afterwards, we present numerical simulation results of bandstructure tuning due to the breaking of 2D hexagonal and square symmetries, and demonstrate formation of Dirac points in liquid lattices with a broken hexagonal symmetry. Finally, we demonstrate that symmetry changes of 2D optical liquid lattices leads to tuning of the corresponding lasing frequencies and the corresponding emission directions. 

\section*{Results}

\subsection*{Light-induced interaction between surface optical modes and a TLD film}

The set of coupled governing equations that describes light-fluid interaction due to TLD film thickness changes includes: Maxwell equations, Navier-Stokes equations for an incompressible Newtonian fluid, balance condition between viscous and surface tension stresses on the gas-fluid (or fluid-fluid) interface, and heat/mass-transport equations depending on the specific light-induced mechanism which triggers local changes of the surface tension. We employ heat-transport and mass-transport equations, relevant for the TC effect and SC effect, respectively, which constitute the coupling mechanism between light propagation and dynamics of TLD film. Specifically, light-induced changes of the surface tension lead to deformation of the TLD film, $\eta(\vec{r}_{\parallel},t)$, which in turn is coupled back to the propagation of light due to the associated spatial changes of the dielectric function, $\Delta \epsilon_{D}$, which in the leading order of $\eta/h_{0}$ is given by \cite{rubin2018nonlocal}
\begin{equation}
	 \Delta \epsilon_{D}(\vec{r}_{\parallel},t) = b \eta(\vec{r}_{\parallel},t)/h_{0}.
\label{DielectricThickness}	 
\end{equation}
Here, $h_{0}$ is the initial thickness of a flat and undeformed TLD film, which we assume to be thinner than the penetration depth of the corresponding optical surface mode into the bulk, and $b$ is the mode-dependent coefficient we discuss below. 
First, we describe the case of a TC driven interaction. 
Applying quasistatic temperature field distribution for the thermal transport and the thin film deformation linear equations,
yields the following nonlocal relation between the deformation $\eta$ and the optical intensity $I \equiv \vert E \vert^{2}$
\begin{equation}
	\eta (\vec{r}_{\parallel},t)/h_{0} = - \dfrac{M}{\tau_{th}}
    \int d\vec{r}^{\prime}_{\parallel} dt^{\prime}  
    G_{l}(\vec{r}_{\parallel}-\vec{r}^{\prime}_{\parallel},t-t^{\prime}) I(\vec{r}^{\prime}_{\parallel},t^{\prime})/I_{0},
\label{GreenGreen22}    
\end{equation}
where $\text{M} \equiv \text{Ma} \cdot \chi/2$ and $G_{l}$ is the corresponding Green's function of the thin film equation (see Supplemental Material of \cite{rubin2018nonlocal}).  
Here, $\text{Ma}=\sigma_{T}\Delta T h_{0}/(\mu D_{th})$ 
is the dimensionless Marangoni number which represents the ratio between the surface tension stresses due to the TC effect and dissipative forces due to fluid viscosity and thermal diffusivity; $\chi \equiv \alpha_{th}^{ }  d^{2} I_{0}/(k_{th}^{ } \Delta T)$ is the dimensionless intensity of the heat source; $I_{0}$ is the characteristic optical intensity; $D_{th}^{ }=k_{th}^{ }/(\rho^{ } c_{p}^{ })$ is the heat diffusion coefficient;  $\rho^{ }$, $c_{p}^{ }$, $k_{th}^{ }$, $\alpha_{th}^{ }$ are the mass density, specific heat,
heat conductance, and optical absorption coefficient, respectively; $\tau_{th}=d^{2}/D_{th}^{ }$ is the typical time scale; $d$ is the typical length scale along the in-plane direction. 
Importantly, Eq.(\ref{DielectricThickness}) and Eq.(\ref{GreenGreen22}) are valid for both SPP and slab WG modes. In particular, the optical absorption coefficient $\alpha_{th}$ which drives thermal interaction between SPPs and TLD film or WG and TLD film, could stem from ohmic losses in metals or from optical absorption in dielectrics, respectively. In fact, optical absorption in dielectrics could be enhanced by mixing the liquid with strong nano-absorbers, as was demonstrated by doping liquid with concentration of
$2–4$ nm diameter CdSe or CdTe nanoparticles of concentration $10^{22}$ m$^{-3}$ (i.e. approximately $0.1$\% of the total liquid volume), triggering TC effect in capillaries
under $514$ nm illumination \cite{lamhot2009optical}, and optical power of approximately $120$ mW concentrated in a region of few mm. In case of TLD film deformation driven by surface optical modes, which admit significantly smaller mode volume (of several hundreds of nm at most), we expect much lower powers needed to generate TLD film deformation. Note, that in TLD film, nanoparticles transport is coupled to both temperature field and to TC flows, which require additional equation in our model; also, local changes of nanoparticles concentration are expected to introduce local changes of shear viscosity \cite{happel1965englewood} which are beyond the scope of this work.

For the case of SC flows, we focus on cis-trans transformation that can be described by the simple phenomenological two-state model $A \xrightleftharpoons[k_{off}]{k_{on}} B$, where $c_{A,B}$ stand for the molar concentration of the corresponding photo-active molecule in the cis/trans states $A$ and $B$, respectively. Assuming linear dependence of the surface tension $\sigma$, with respect to small changes of $c_{A,B}$, and furthermore assuming for simplicity that both species admit identical molecular diffusion coefficients $D$ and similar affinity to the gas-fluid interface,
leads to the following nonlocal relation between the deformation $\eta$ and the optical power $I$, which is similar to the relation Eq.(\ref{GreenGreen22}) relevant to the TC case discussed above (see Supplemental Material for derivation)
\begin{equation}
	\eta (\vec{r}_{\parallel},t)/h_{0} = - \dfrac{\text{M}_{c} }{\tau_{on}} 
    \int d\vec{r}^{\prime}_{\parallel} dt^{\prime} 
    G_{l}(\vec{r}_{\parallel}-\vec{r}^{\prime}_{\parallel},t-t^{\prime}) I(\vec{r}^{\prime}_{\parallel},t^{\prime})/I_{0}.
\label{GreenGreenSC}
\end{equation}
where $\text{M}_{c} \equiv \text{Ma}_{c} \cdot c_{0}/\Delta c$.
Here, $\text{Ma}_{c} \equiv h_{0}  \overline{\sigma}_{c} \Delta c /(\mu D)$ is the
dimensionless SC Marangoni number which represents the ratio between the surface tension stresses due to the SC effect, and dissipative forces due to fluid viscosity and molecular diffusivity; $\tau_{c}=d^{2}/D$ is the typical diffusion time and $\tau_{on}=1/k_{on}$ is the typical transformation rate for a given optical intensity (see Supplemental Material). 

Employing a perturbative expansion for the SPP and slab WG modes (see Supplemental Material), 
which incorporates the effects of diffraction and gain, yields the following complex nonlocal Ginzburg-Landau (GL) equation (see Supplemental Material for derivation)
\begin{equation}
\begin{split}
	&2 i \beta_{0} \dfrac{\partial A}{\partial z} + \dfrac{\partial^{2} A}{\partial y^{2}} - V[A(\vec{r}_{\parallel}) ] A + i \Gamma A = 0, 
\\
	V [A(\vec{r}_{\parallel}) ] \equiv &- \Delta \epsilon_{d} [A(\vec{r}_{\parallel}) ]  =- \chi_{TC/SC} \int d\vec{r}_{\parallel}^{\prime} G_{l}(\vec{r}_{\parallel}^{ },\vec{r}_{\parallel}^{\prime}) \vert A^{ } (\vec{r}_{\parallel}^{\prime}) \vert^{2}.
\end{split}	
\label{GinzburgLandau}
\end{equation}
A complex GL equation with a local potential was shown to capture the effects of nonlinear Kerr media (see \cite{aranson2002world} and references within), propagating SPPs adjacent to solid dielectric with gain \cite{marini2010ginzburg}, as well as nanofocusing of SPPs in tapered plasmonic waveguides \cite{davoyan2010nonlinear}.
The nonlocal integral term, $V[A(\vec{r}_{\parallel})]$, in Eq.\ref{GinzburgLandau} represents the self-induced potential due to refractive index changes which stem from liquid deformation extending beyond the region of maximal optical intensity. For the case of a TLD film interacting with an SPP mode, the constant $\chi_{TC/SC}= f b \text{M}$ is a product of three dimensionless numbers incorporating the effects of SPP enhancement, depth averaging and TC or SC effects, whereas for the case of a TLD film interacting with slab WG modes the constant is $\chi_{TC/SC}= b \text{M}$.
Importantly, the case where a TLD film introduces nonlocal loss or gain, is incorporated through a complex constant $b$, where its' positive/negative imaginary part correspond to effects of loss/gain (i.e. damped/amplified modes), respectively. 
The term, $i \Gamma$ on the other hand represents local loss or gain, which are not related to refractive index changes due to TLD film deformation. For instance, local change of the dielectric function due to gain changes is given by $i \Gamma = \epsilon_{m}^{2} \delta \epsilon_{d}/(\epsilon_{d0}+\epsilon_{m})^{2}$
where $\delta \epsilon_{d}$ is the optically-induced dielectric function change in the gain media. In dye solutions, the latter may stem either from intensity dependent modulations, which affect the population of the electronic states of the dye molecules, or from temperature gradients formed by the heat generated by relaxation of the excited molecules (see \cite{eichler2013laser} and references within). 

\subsection*{1D and 2D optical liquid lattices with tunable symmetry}

\begin{figure*}
	  \includegraphics[width=\textwidth]{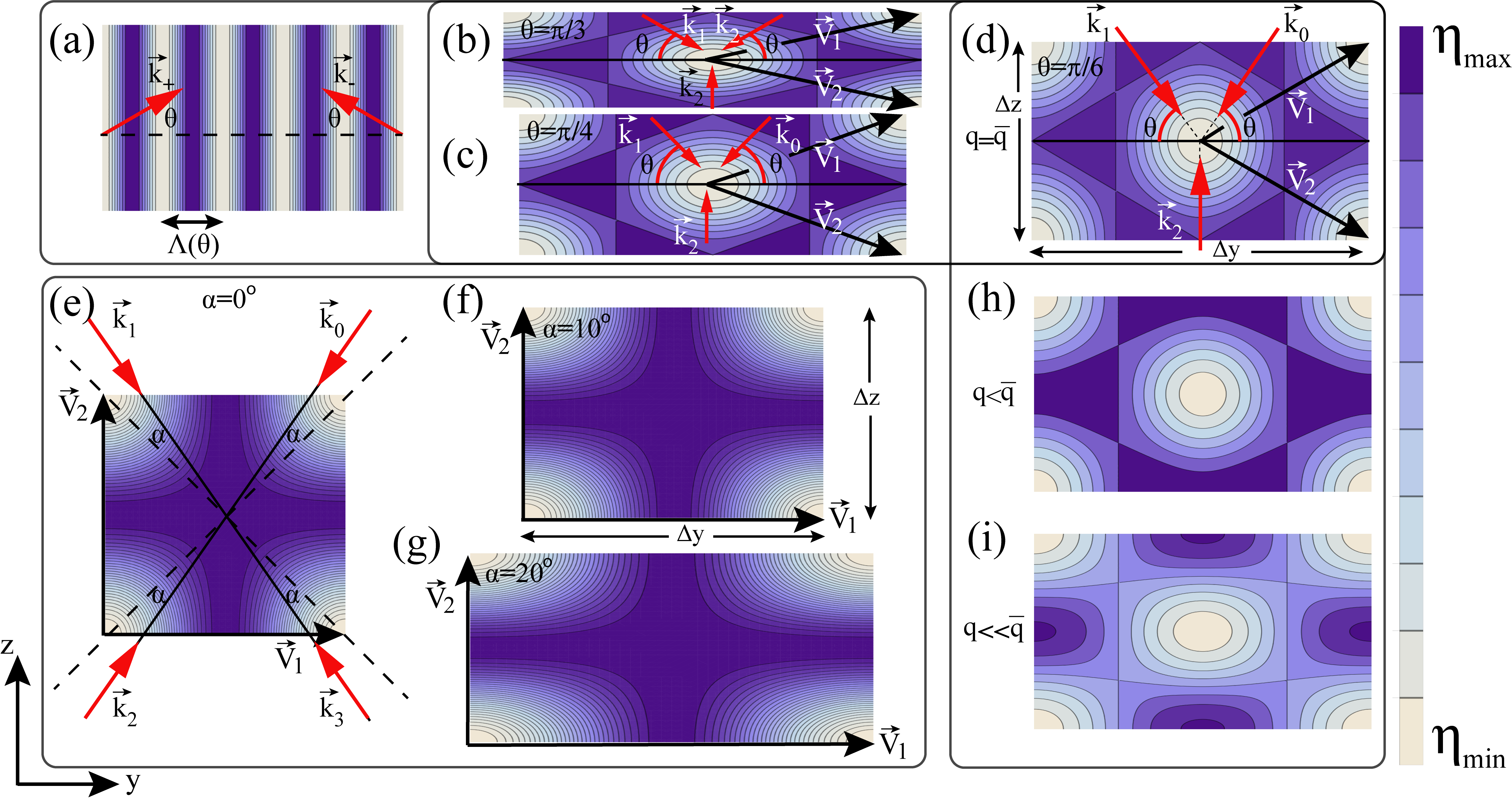}
    \caption{Optical liquid lattices formed from a TLD film due to interference of SPPs or WG plane wave modes with propagating direction marked with red arrows. (a) 1D lattice formed by interference of two surface optical plane waves with $\theta$ dependent periodicity given by Eq.(\ref{PeriodicDeformation}); 2D lattices of hexagonal symmetry (b-d) and rectangular symmetry (e-g) where $\vec{V}_{1}$ and $\vec{V}_{2}$ are the corresponding primitive vectors. The  symmetry of the liquid lattice and can be controlled by the relative angles of the interfering beams; (b) $\theta=30^{o}$ and (c) $\theta=45^{o}$ are lattices with hexagonal broken/distorted symmetry whereas (d) $\theta=60^{o}$ is hexagonal symmetric lattice; (e) $\alpha=0^{0}$ (square symmetry), (f) $\alpha=10^{o}$ and (g) $\alpha=20^{o}$ correspond to rectangular symmetry. (h,i) Optical liquid lattice for different values of $q$ which results in a phase transition from a hexagonal lattice (d) to merging of the triangular sites (h) and faced centered cubic lattice (i). The 2D liquid lattices are normalized by $\eta_{max}$ and $\eta_{min} \equiv -\eta_{max}$ defined in Eq.(\ref{EtaMaxHex}) and below Eq.(\ref{SquarePeriod}).}
    \label{Interference}
\end{figure*}
Next we show that by controlling the interference pattern of surface optical modes, it is possible to form optical liquid lattices directly from a TLD film, and tune their symmetries.
Consider an interference of $N$ plane waves of real-valued amplitudes $a_{n}$, that propagate in the $y-z$ plane along the directions formed by angles $\theta_{n}$, which are measured relative to the positive direction of the $y$-axis. The corresponding wave vectors are given by $\vec{k}_{n}=k_{0}\left(\cos(\theta_{n}), \sin(\theta_{n}) \right)$, where $k_{0}=2 \pi /\lambda$ and $\lambda$ is the effective wavelength which is set by the relevant frequency and the corresponding dispersion relation; i.e. for SPP or WG mode the wavelength is given by $\lambda = \lambda_{0}/n$ where $\lambda_{0}$ is vacuum wavelength and $n$ is the corresponding effective refractive index. The resultant optical intensity $I$, due to interference of the plane waves $a_{n}e^{i \vec{k}_{n} \cdot \vec{r}_{\parallel}}$ is then given by
\begin{equation}
\begin{split}
	&I = \Big \vert \sum_{n=0}^{N-1} a_{n} e^{i  \vec{k}_{n} \cdot \vec{r}_{\parallel} }  \Big \vert^{2} = 
\\	
	&I_{N} + 2\sum \limits_{n \neq m}^{N-1} a_{n}a_{m} \cos \left( (\vec{k}_{n}-\vec{k}_{m}) \cdot \vec{r}_{\parallel} \right); \quad I_{N} \equiv \sum_{n=0}^{N-1} \vert a_{n} \vert^{2}.
\label{Intensity}
\end{split}	
\end{equation}
In this work we focus on the cases $N=2,3,4$, which as shown shortly below correspond to optical liquid lattices with 1D translation, 2D hexagonal and 2D rectangular symmetries, respectively. Furthermore, we allow non-equal relative propagation angles between adjacent wave vectors and non-equal amplitudes, leading to broken symmetries and to changes of the symmetry group of the formed optical liquid lattices.

First, consider the simplest $N=2$ case, that leads to a 1D optical liquid lattice. The intensity distribution of two optical plane waves  of equal intensity $I_{0}$ propagating along the $y-z$ plane with wave vectors $\vec{k}_{\pm} = k (\pm \cos (\theta), \sin(\theta)$ (see Fig.\ref{Interference}(a)), is given by $2I_{0}(1+\cos(K(\theta) x))$,  which upon inserting into 
Eq.(\ref{GreenGreen22}) yields the corresponding 1D deformation of the TLD film 
\begin{equation}
\begin{split}
	&\dfrac{\eta}{h_{0}}  = - \dfrac{2 \text{M}}{\tau_{th} \lambda_{\bar{n}}} \cos(K(\theta) x); 
\\	
	\quad K(\theta) = & 2 k \cos(\theta); \quad \Lambda (\theta) \equiv \dfrac{2 \pi}{K(\theta)} = \dfrac{\pi}{k \cos(\theta)},
\label{PeriodicDeformation}
\end{split}
\end{equation}
where $\lambda_{\bar{n}}$ is a constant (see Supplemental Material).
Importantly, the amplitude of the deformation is proportional to the Marangoni constant, and the periodicity $\Lambda(\theta)$ can be tuned by the angle $\theta$. Employing the definitions for $\text{M}, \chi$ and $\tau_{th}$ near Eq.(\ref{GreenGreen22}), the modulation amplitude of the cosine deformation described in Eq.(\ref{PeriodicDeformation}) can be written as $2 \text{M}/(\tau_{th} \lambda_{\bar{n}}) = (3 \sigma_{T} \alpha_{th} d_{x}^{4}/(\sigma_{0} k_{th} h_{0}^{2} \pi^{4})) I_{0}$. Inserting the following typical parameters that can be satisfied by silicone oil on gold substrate: $\sigma_{0}=10^{-3}$ Nm$^{-1}$, $\sigma_{T}=10^{-3}$ Nm$^{-1}$K$^{-1}$, $\alpha_{th} = 7.7 \cdot 10^{7}$ m$^{-1}$, $k_{th} = 3 \cdot 10^{2}$ Wm$^{-1}$K$^{-1}$, $d_{x} = 50$ $\mu$m, $h_{0}= 450$ nm, $\bar{n}=20$; we learn that $2 \text{M}/(\tau_{th} \lambda_{\bar{n}}) = 1.5 \cdot 10^{-7} I_{0}$. 
Consequently, using Eq.(\ref{PeriodicDeformation}) the optical intensity required to generate periodic deformation $\eta$ of amplitude $450$ nm which is similar to TLD film thickness (i.e. $\eta/h_{0} \approx 1$), is of the order of magnitude $I_{0} = 6.5 \cdot 10^{6}$ Wm$^{-2}$.

Next we consider the case $N=3$, which leads to a 2D symmetric liquid lattice, where the relative amplitudes of the three beams satisfy $a_{1}=a_{2}=a_{3}/q \equiv a$ where $a$ and $q$ are real constants, 
and assume the corresponding propagation directions are given by $\hat{n}_{1}= (\cos(\theta), -\sin(\theta))$, $\hat{n}_{2}= - (\cos(\theta), \sin(\theta))$, $\hat{n}_{3}=(0,1)$ (which can be also written as $\theta_{n} =(n+1) \pi + (-1)^{n} \theta $ for $n=0,1$, and $\theta_{2}=\pi/2$).
Inserting the corresponding intensity distribution (see Supplemental Material) into Eq.(\ref{GreenGreen22})
yields the following TLD film deformation
\begin{equation}
\begin{split}
 	\dfrac{\eta(\vec{r}_{\parallel},t)}{h_{0}} =- \dfrac{\eta_{max}}{3} \Big[ \cos \left( \dfrac{4 \pi y}{\Delta y} \right) + 	
 	2\cos \left( \dfrac{2 \pi y}{\Delta y} \right) \cos \left( \dfrac{2 \pi z}{\Delta z} \right)
 	 \Big]. 
\end{split}
\label{HexDeformation}	  
\end{equation}
Here, $\Delta y$ and $\Delta z$ are the corresponding periodicities along the $y$ and $z$ directions, respectively, 
\begin{equation}
	\Delta y = \dfrac{\lambda}{\cos(\theta)}; \quad
	\Delta z = \dfrac{\lambda}{1+\sin(\theta)},
\label{HexIntensityPer}	
\end{equation}
which are set by the angle $\theta$ and the wavelength of the optical surface wave $\lambda$ but do not depend on the amplitude $a$ nor on $q$, and
$\eta_{max}$ is a constant given by
\begin{equation}
	\eta_{max} \equiv (3\bar{q} \text{M})/(\lambda_{\bar{n},0} \tau_{th}); \quad \bar{q}=2\lambda_{2\bar{n},0}/\lambda_{\bar{m},\bar{n}},
\label{EtaMaxHex}		
\end{equation}
where $\bar{m},\bar{n}$ are integers that satisfy $\bar{m}=2 d_{y}/\Delta y$, $\bar{n}=2 d_{z}/\Delta z$ (see Supplemental Material for more details). 

Expanding the deformation given by Eq.(\ref{HexDeformation}) up to a second order in Taylor series around one of the maximum points (e.g. $(\Delta y/3,0)$), yields a harmonic oscillator potential
which admits a degree of anisotropy given by $3 (\Delta z/ \Delta y)^{2}$, which is a monotonic decreasing function of $\theta$. 
In particular, the case $\theta = \pi/6$, i.e. when the relative angles between the interfering plane waves all equal to $2 \pi/3$, the degree of anistropy equals to unity and the fluid deformation admits a hexagonal symmetry. 
Values of $\theta$ which differ from $\pi/6$ result in a fluid deformation with a distorted unit cell and of broken hexagonal symmetry. Interestingly, the unit cell area $\Delta y \cdot \Delta z/2$, is not preserved under changes of the interfering plane-waves' relative angle, and the minimal unit cell area corresponds to $\theta=\pi/6$. Fig.\ref{Interference}(b-d) present fluid deformation described by Eq.(\ref{HexDeformation}) for the values $\theta=\pi/6$, $\theta=\pi/4$ and $\theta=\pi/3$, respectively.  
Other values of the parameter $q$ which are close to $q=\bar{q}$ can be used to control fluid distribution within each unit cell. Values of $q$ which differ significantly from $q=\bar{q}$ can be employed to trigger phase transition of the fluid lattice. 
In particular, increasingly smaller values of $q$ (i.e. larger $a_{1,2}/a_{3}$) and fixed propagation directions which do not modify $\Delta y$ and $\Delta z$, lead to change of the fluid deformation symmetry from hexagonal, which is composed of two equivalent inter-penetrating triangular Bravais lattices which admit two primitive vectors $\vec{V}_{1,2}$ and and two cites per unit cell \cite{wallace1947}, to a face centered cubic symmetry with one primitive vector per unit cell. Fig.\ref{Interference}(d,h,i) present TLD film deformation for the case $\theta=\pi/6$ and successively smaller values of $q$ leading to merging of fluid peaks (blue) within each cell and a phase transition from hexagonal lattice to a lattice with centered cubic symmetry.

A liquid lattice with a rectangular tunable symmetry can be formed by interfering four beams where the relative amplitudes satisfy $a_{0,2}=q a$, $a_{1,3}=a$ and the propagation angles are given by $\theta_{n}=(1+2n) \pi/4+(-1)^{n}\alpha$ ($n=0,1,2,3$), where $\alpha$ is a real parameter. 
Inserting the corresponding intensity (see Supplemental Information) into Eq.(\ref{GreenGreen22}), yields the following TC-driven deformation of the TLD film
\begin{equation}
 \begin{split}
 	\dfrac{\eta(\vec{r}_{\parallel},t)}{h_{0}} = - \dfrac{\eta_{max}}{3} \Big[&\cos \left( \dfrac{2 \pi y}{\Delta y} \right) + 
 	\cos \left(\dfrac{2 \pi z}{\Delta z}\right) +
 \\		
 	 &\cos \left( \dfrac{2 \pi y}{\Delta y} \right) \cos \left( \dfrac{2 \pi z}{\Delta z} \right)
 	  \Big].
 \end{split}
 \label{SquareDeformation}	  
\end{equation}
Here, $\Delta y$ and $\Delta z$ are the corresponding periodicities along the $y,z$ directions, given by
\begin{equation}
 	\Delta y = \dfrac{\lambda}{2 \left( \cos(\alpha) - \sin(\alpha) \right)}; \quad \Delta z = \dfrac{\lambda}{2 \left( \cos(\alpha) + \sin(\alpha) \right)};
 \label{SquarePeriod}	
\end{equation}
and $\eta_{max}$ is a constant given by
 $\eta_{max} \equiv 3\text{M}(\bar{q}-3/4)/(2\lambda_{\bar{n},0} \tau_{th})$ where $\bar{m}, \bar{n}$ are integers that satisfy $\bar{m}=2 d_{y}/\Delta y$ and $\bar{n}=2 d_{z}/\Delta z$, and the constant $\bar{q}$ satisfies the quadratic equation $(1/2-\bar{q}^{2}/4)/(2\lambda_{\bar{m},\bar{n}})=(\bar{q}-3/4)/\lambda_{\bar{n},0}$.
Since $\lambda_{\bar{m},\bar{n}}$ and $\lambda_{\bar{n},0}$ are both positive, the latter quadratic equation is guaranteed to admit two different and positive solutions for $\bar{q}$. Fig.\ref{Interference}(d-f) presents a TLD film deformation described by Eq.(\ref{SquareDeformation}) for increasingly higher  values of $\alpha$ ($\alpha=0^{o}, 10^{o}, 20^{o}$), leading to an increase of $\Delta y$ and to a decrease of $\Delta z$ as follows from Eq.(\ref{SquarePeriod}). Similarly, the case of SC-driven deformation of TLD film is obtained by substituting the corresponding intensity into Eq.(\ref{GreenGreenSC}), leading to identical expressions for the deformations given by Eq.(\ref{HexDeformation}) and Eq.(\ref{SquareDeformation}), where the ratio $\text{M}/\tau_{th}$ in $\eta_{max}$ is replaced by $\text{M}_{c}c_{b}/\tau_{c}\Delta c$.

\subsection*{Self-induced distributed feedback lasing in 1D optical liquid lattices}

The self-induced 1D deformation of the TLD film given by Eq.(\ref{PeriodicDeformation}) (see also Fig.\ref{Interference}(a)), leads to a periodic modulation of the depth-averaged refractive index for the corresponding surface optical modes which triggers a self-induced selection mechanism. The latter singles out a set of potential lasing modes that satisfy the minimum phase, or the $q$-th order Bragg resonant condition, given by 
\begin{equation}
	2 \Delta \equiv k_{+}^{L}- k_{-}^{L} - q K =0,
\label{ResCondition}	
\end{equation}	
where $\Delta$ is the mismatch parameter \cite{lifante2003integrated,suhara2004semiconductor} between the wave numbers $k_{\pm}^{L}$ of the lasing modes and $K$ is defined by Eq.(\ref{PeriodicDeformation}). The case where the modes admit equal wave vectors and satisfy $k_{+}^{L} = - k_{-}^{L} \equiv k_{L}$,
leads to a lasing mode wave vector of magnitude $k_{L} =q k\cos(\theta)$ and wavelength $\lambda_{L} = \lambda/(q\cos(\theta))$, where the integer $q$ is determined by the spectral region of maximal spontaneous emission of the gain media and $\theta$ is the tunable angle of the interfering surface opitcal modes schematically described in Fig.\ref{Interference}(a). For instance, in a case of two counter-propagating ($\theta=0$) interfering TE WG modes of wavelength $1.55$ $\mu$m, 
the resonant condition given by Eq.(\ref{ResCondition}) leads to $\lambda_{L}=775$ nm and $q=2$. 
For the case $\theta=\pi/3$, identical resonant wavelength $\lambda_{L}=775$ nm is realized for $q=4$.
Note that the effective mode index, $n_{eff}$, defined via the relation $k = n_{eff} k_{0}$ cancels from both sides of the resonant condition, because in the self-induced case the pumping beam coincides with the beam that forms the grating. 
In the limit case of non-self-induced lasing regime, i.e., in a case the pair of interfering surface optical plane waves with wave vectors of equal magnitude $k$ are used to form and tune the 1D liquid lattice of periodicity $\Lambda=\pi/(k \cos(\theta))$ whereas the pumping is performed by an additional unfocused incident source of free-space wavelength $\lambda_{p}$ (e.g. similarly to the method employed in \cite{zhang2006mechanically}), the lasing condition takes the form $2 n_{eff} \Lambda -q \lambda_{p}= 0$. 
For instance, a $300$ nm thick suspended liquid film of index $1.409$, the corresponding $n_{eff}$ of the TE mode is given by $n_{eff} = 1.319$, and increasing $\theta$ from $0$ to $\pi/6$, modifies the lattice period $\Lambda$ from $1$ $\mu$m to $2$ $\mu$m and consequently shifts the resonance condition from $q=5$ and $\lambda_{p}=527.6$ nm to $q=7$ and $\lambda_{p}=753.7$ nm.

Note, that similarly to other self-induced lasing mechanisms \cite{kogelnik1972coupled}
the field intensity plays a double role; it determines both the coupling constant \cite{yariv2006photonics, liu2009photonic} of the backward Bragg scattering (which provides the feedback mechanism) as well as the pump intensity necessary for lasing. Nevertheless, in our case, the amplitude of the optically induced periodic liquid lattice and consequently the coupling between the right- and the left-propagating modes is a more pronounced function of the optical intensity due to the higher difference between the typical refractive indices of liquids and gases. Notably, the relatively low optical power needed to introduce index changes induced by
the TC/SC gas-liquid interface deformation (see \cite{rubin2018nonlocal}, below eq.13 for the TC case), is expected to lead to substantially lower lasing threshold as compared to a thick liquid film or a liquid without gas-liquid interface (e.g., liquid fully occupying a microchannel), where the liquid-light interaction due to interface deformation is expected to be absent.
In particular, in the absence of gas-liquid interface, index changes in liquids mostly stem from changes of polarizability, population of the electronic states (see \cite{eichler2013laser} and references within) and material density; for instance the index change due to thermo-optical effect, $\Delta n_{TO}$, is given by $\Delta n_{TO} = \alpha_{TO} \Delta T$ where $\Delta T$ is a temperature change $\alpha_{TO}$ is the thermo-optical coefficient which is typically of the following order of magnitude $\alpha_{TO} = 10^{-4}$ K$^{-1}$. In case of thermocapillary effect, the corresponding index change driven by identical temperature difference is given by $\Delta n_{TC} = \alpha_{TC} \Delta T$, where 
\begin{equation}
	\alpha_{TC} =  \dfrac{3 \sigma_{T} d_{z}^{2}}{2 \sigma_{0} h_{0}^{2} \pi^{2}} \dfrac{(n-1/2)^{2}}{n^{4}}
\end{equation}	 
and $n$ is integer indicating the numbers of optical periods that fit into the slot (see Supplementary Material for derivation). Inserting the typical numbers used below Eq.(\ref{PeriodicDeformation}) and taking $n=20$, we learn that $\alpha_{TC} = 187.6 $. The latter indicates that under similar temperature change, the amplitude of the periodically induced index and consequently also the coupling coefficient between the counter propagating modes (which directly determines the lasing threshold) is significantly higher for the thermocapillary case.


In the limit of small deformation and small intensity the corresponding coupling constant can be written as $\kappa = \kappa_{0} I_{0}$, where $\kappa_{0}$ is the constant of proportionality given by
\begin{equation}
	\kappa_{0} = k \overline{\text{M}} h_{0}; \quad \overline{\text{M}} =\dfrac{2 \text{Ma}}{\tau_{th} \lambda_{N}} \cdot 			\dfrac{\alpha_{th}^{ }  d^{2}}{k_{th}^{ } \Delta T}.
\end{equation}
Here, $h_{0}$ is the mean thickness of the TLD film, and $k$ is a mode-dependent coefficient; e.g., for WG TE-TE coupling, the expression for $k$ is given in \cite{liu2009photonic} (see Supplemental Material).
The coupling coefficient affects the oscillation condition
and for DFB laser cavities it can be written as the following complex equation \cite{kogelnik1972coupled} 
\begin{equation}
	\kappa L \sinh(i \gamma L) = \pm \gamma L; \quad  \gamma = -i \Big[ \left(\dfrac{g}{2} + i \Delta \right)^{2} + \kappa^{2} \Big]^{1/2},
\label{DFBoscillationcond}	 
\end{equation}
where the corresponding solution for $g$ and $\Delta$ must satisfy simultaneously vanishing of the real and the imaginary parts of Eq.(\ref{DFBoscillationcond}). While index modulation typically supports lasing at the edges of Brillouin zone, the case of a gain coupling which is described by Eq.(\ref{DFBoscillationcond}) with $\kappa \rightarrow i \kappa$, results in lasing at the center of the Brillouin zone. In our study both can be realized depending on the dielectric properties of the TLD film. Fig.S3 presents the solution of Eq.(\ref{DFBoscillationcond}) in the index and gain modulation regimes, as a function of increased intensity leading to decreasing of the lasing threshold.

\subsection*{Bandstructure of WG and SPP modes in hexagonal and rectangular liquid lattices}

\begin{figure*}
	  \includegraphics[width=\textwidth]{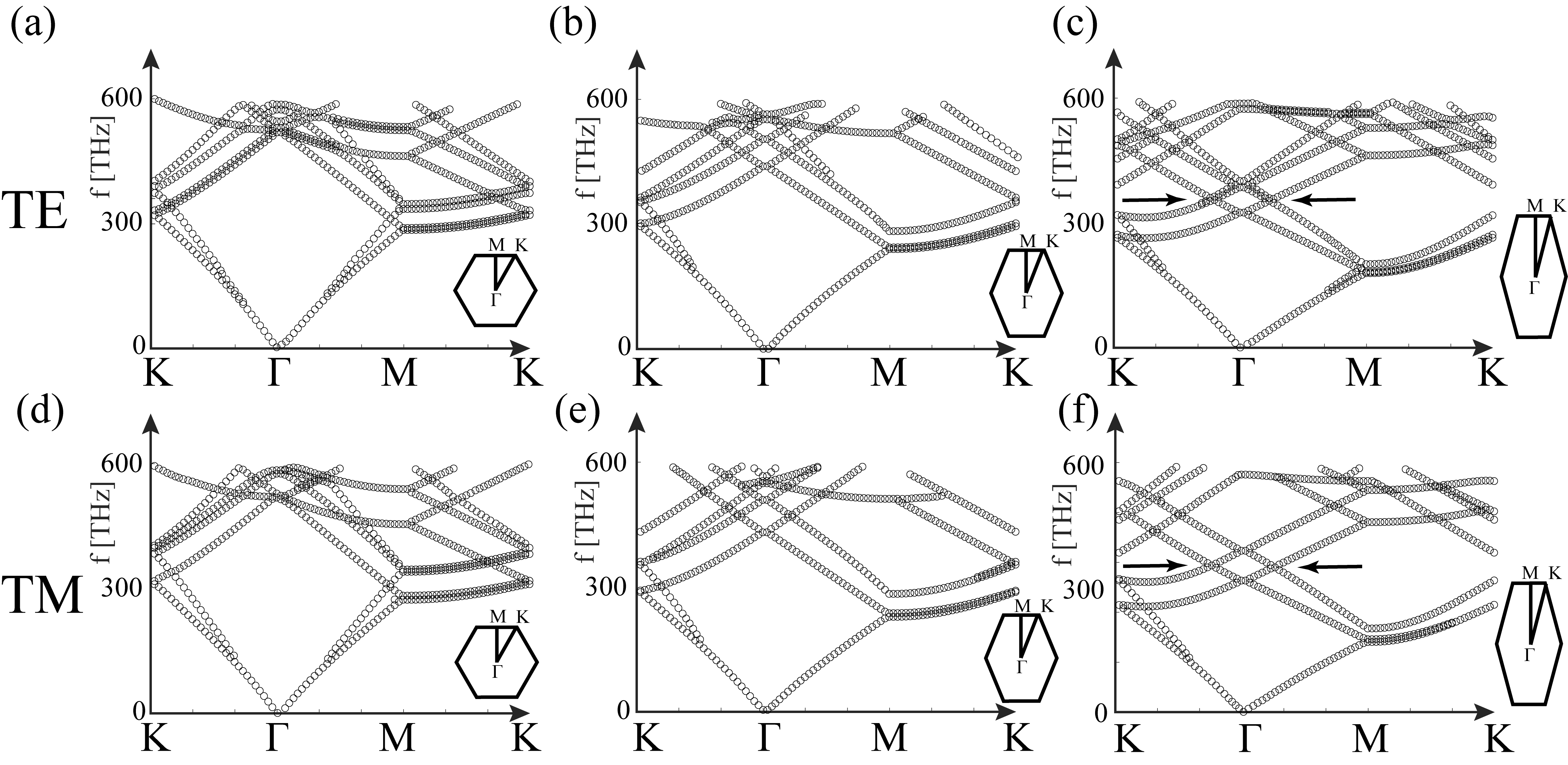}
    \caption{Numerical simulation results presenting the bandstructure of photonic slab WG modes (i.e. frequency as a function of the corresponding directions in the $k-$space) formed in a suspended photonic liquid lattice of hexagonal symmetry and hexagonal broken symmetry. (a-c) Bandstructure of TE slab WG modes in a suspended photonic liquid lattice presented in Fig.\ref{Setup}(b-d) due to interference formed by three plane waves at angles $\theta = \pi/6$ (a), $\pi/4$ (b) and $\pi/3$ (c). (d-f) Bandstructure of TM slab WG modes at interference angles $\theta = \pi/6$ (d), $\pi/4$ (e) and $\pi/3$ (f). Two Dirac points emerge at the angle $\theta=\pi/3$ at frequencies around $353$ THz and $349$ THz, respectively, marked at (c) and (f) by black arrows. In the simulation, the refractive index is $1.409$, the mean TLD film thickness is $450$ nm and the peak to peak undulation amplitude is $300$ nm.}
    \label{HexagonsWG}
\end{figure*}

Consider the case of interferring plane waves of the type presented in Eq.(\ref{Intensity}), which interact via nonlocal and nonlinear integral term in Eq.(\ref{GinzburgLandau}). Representing the intensity distribution given by Eq.(\ref{Intensity}) as a linear superposition of elements in the $\varphi_{m,n}$ basis and employing orthogonality of these basis functions, leads to the following dispersion relation 
\begin{equation}
\begin{split}
	(2 i \beta_{0} &k_{z}^{(i)} + (k_{y}^{(i)})^{2})a_{i} = 4\chi_{TC/SC} \sum\limits_{m,n=0}^{N-1} \Delta_{imn} \dfrac{a_{i} a_{m} a_{n}}				{\lambda_{\bar{m},\bar{n}}}; 
\\	
	&\Delta_{imn} \equiv \dfrac{1}{d_{y}d_{z}} \int\limits_{0}^{d_{y}}\int\limits_{0}^{d_{z}} e^{-i \vec{k}_{\parallel}^{(i)} \cdot \vec{r}_{\parallel}} 			\varphi_{m,n}(\vec{r}_{\parallel}) d \vec{r}_{\parallel},
\label{CGLEplaneWave}
\end{split}	
\end{equation}
where the nonlocal term is reduced to a cubic nonlinear term and $\bar{n},\bar{m}$ satisfy $k^{(m)}_{y}-k^{(n)}_{y} = \bar{m} \pi/d_{y}$ and $k^{(m)}_{z}-k^{(n)}_{z} = \bar{n} \pi/d_{z}$. The system of equations given by Eq.(\ref{CGLEplaneWave}) can be solved perturbatively, and in the lowest order of perturbative expansion in the dimensionless parameter $\chi_{TC/SC}$, which corresponds to the physical case of low energetic propagating SPP/WG mode, the corresponding governing complex GL equation given by Eq.(\ref{GinzburgLandau}), simplifies to a linear Schr\"{o}dinger equation given by
\begin{equation}
\begin{split}
	2 i \beta_{0} \dfrac{\partial A}{\partial z} &+ \dfrac{\partial^{2} A}{\partial y^{2}} - V^{(0)}(\vec{r}_{\parallel}) A = 0; 
	\\
	 V^{(0)}(\vec{r}_{\parallel}) \equiv - \Delta \epsilon_{d} &=- \chi_{TC/SC} \int d\vec{r}_{\parallel}^{\prime} G_{l}(\vec{r}_{\parallel},\vec{r}_{\parallel}^{\prime}) \vert A^{(0)} (\vec{r}_{\parallel}^{\prime}) \vert^{2}.
\end{split}
\end{equation}
Here, $V^{(0)}(\vec{r}_{\parallel}) = - \chi_{TC/SC}  a_{m}^{(0)} a_{n}^{(0)} \varphi_{m,n}(\vec{r}_{\parallel}) / \lambda_{m,n}$ is the self-induced potential function and $a^{(0)}$ are the lowest order non-perturbed envelope functions.
For the specific case of three interfering beams with propagation directions specified following Eq.(\ref{Intensity}) 
and amplitudes that satisfy the condition specified following Eq.(\ref{HexDeformation}), the TLD film deformation is given by Eq.(\ref{HexDeformation}) and the corresponding induced optical potential for TC case is given by
\begin{equation}
\begin{split}
	V^{(0)}(\vec{r}_{\parallel})  = &V_{0} \left( \cos(2 \bar{n} \beta x) + 2 \cos( \bar{n} \beta x) \cos( \bar{m} \beta y) \right); 
\\	
	&V_{0} \equiv \bar{q} b \text{M}/(\lambda_{\bar{n},0} \tau_{th}) = b \eta_{max}/3,
\end{split}	
\label{HexPotential}	 
\end{equation} 
where the constant $V_{0}$ depends on the optical intensity and the Marangoni constant.
In particular, for a positive Marangoni constant $\sigma_{T}>0$ ($\text{M}>0$) and $b>0$ the TLD film deformation given by Eq.(\ref{HexDeformation}) admits peak values of TLD film deformation in a triangular lattice and minimum values in a hexagonal lattice, and vice versa for the induced potential $V_{0}(\vec{r}_{\parallel})$. 
The opposite case of $\text{M}<0$, which could be realized either by $\sigma_{T}<0$ and $b>0$, or by $\sigma_{T}>0$ and $b<0$, yields an opposite behavior, i.e. liquid accumulation and potential minima around triangular lattice points, and liquid wells and potential maxima around hexagonal lattice points. 

\begin{figure*}
	\includegraphics[width=\textwidth]{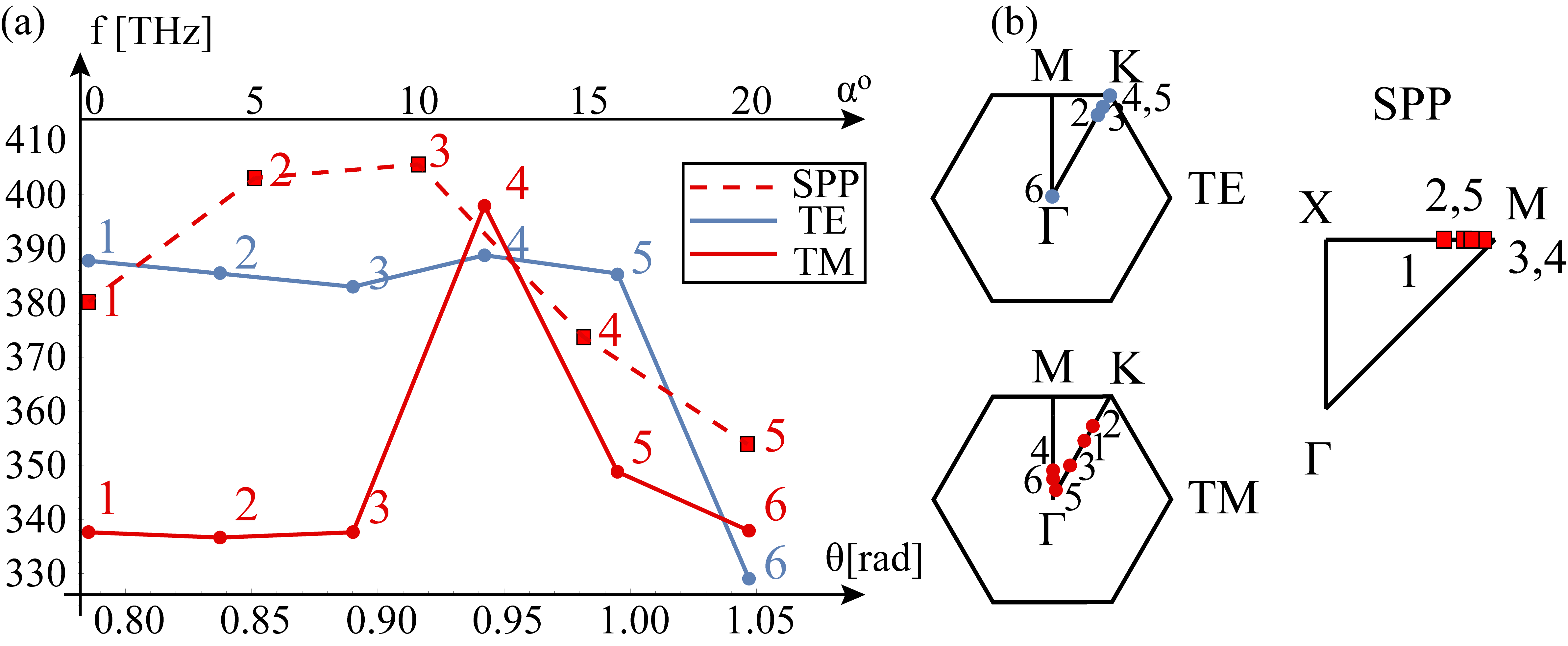}
    \caption{Numerical simulation results presenting lasing frequencies tuning in plasmonic liquid lattice and suspended photonic liquid lattice, of hexagonal and rectangular symmetries, respectively, where the angles of the interfering ``writing beams'' $\alpha$ and $\theta$ are described in Fig.\ref{Interference}. (a) Lasing frequencies tuning of WG TE and TM modes for hexagonal symmetric photonic liquid lattice with $\theta$ values $\theta=45^{o}$, $48^{o}$, $51^{o}$, $54^{o}$, $57^{o}$, $60^{o}$ (in radians) labeled as $1,2,3,4,5,6$ near the blue and red disk, respectively; SPP modes in rectangular symmetric plasmonic liquid lattice with $\alpha$ values $\alpha=0^{o}$, $5^{o}$, $10^{o}$, $15^{o}$, $20^{o}$, labeled respectively as $1,2,3,4,5$ near the red squares. (b) The location of the corresponding lasing modes in the reduced Brillouin zone in the k-space.}
    \label{LasingFreq}
\end{figure*}

To get an insight into light propagation in a hexagonal crystal one usually adopts the tight binding approximation \cite{ablowitz2010evolution}, 
which describes the dynamics of light as hopping between nearest lattice sites. This approximation is applicable to cases 
when the potential well at each site is sufficiently deep, leading to localized intensity of the Bloch modes around the sites. 
For instance, one could in principle expand the potential Eq.(\ref{HexPotential}) around the peak up to second order \cite{ablowitz2010evolution}, and then approximate it by the exponential function, to obtain closed form expressions for the orbitals as a function of the constant $V_{0}$ (see Supplemental Material). 
Since the 2D hexagonal fluidic lattice admits two sites per one unit cell, the corresponding Wannier function is a linear combination of the two associated orbitals. By taking the continuous limit of the discrete system \cite{ablowitz2010evolution}, the Schr\"{o}dinger equation around high symmetry points in the k-space can be represented as a pair of Dirac equations \cite{ablowitz2013nonlinear}.
The latter implies the presence of a Dirac point around the corresponding high symmetry points (which usually appears around K points), though may also emerge in the inner regions of the Brillouin zone. 

To determine the dispersion relation for SPP or slab WG modes in the regime that is not satisfied by the tight binding approximation and captures the effect of broken hexagonal and cubic symmetries, we turn to numerical simulations. Fig.\ref{HexagonsWG} presents simulation results of the bandstructure for  three interfering beams, with angles $\theta = \pi/6$, $\pi/4$ and $\pi/3$, that lead to a formation of the suspended TLD film configuration (presented in Fig.\ref{Setup}(b)) with fluid topography presented in Fig.\ref{Interference}(a-c), respectively.
Interestingly, optical liquid lattice with increasingly higher broken hexagonal symmetry due to modified interference pattern, shows emergence of Dirac points for the case $\theta=\pi/4$ and $\theta=\pi/3$, i.e. regions with linear dispersion relation, which do not appear in the case $\theta=\pi/6$. The latter shows formation of two Dirac points (marked with black arrows in Fig.\ref{HexagonsWG}(c,f)), and admit the values $352.3$ THz and $354.3$ THz for TE mode, and $350.9$ THz and $347.4$ THz for the TM mode. 
Notably, while Dirac cones are known to emerge in all-dielectric, solid-made and non-reconfigurable setups, in our setup they emerge in reconfigurable optical liquid lattices. In particular, taking advantage of the compliant property of the liquid film allows position tuning of the Dirac points under optically induced surface tension stresses, which is markedly different than other recently suggested mechanisms to control topological properties of light in liquids, such as electrical tunability of liquid crystals \cite{shalaev2018reconfigurable}.

\subsection*{Self-induced distributed feedback lasing and tuning in 2D liquid lattices}

Fig.\ref{LasingFreq} presents simulation results of the lasing frequencies and the corresponding wave vectors in the k-space, within plasmonic and suspended photonic liquid lattices of tunable symmetry formed by interfering surface optical waves described by angles $\theta$ and $\alpha$, described in Fig.\ref{Interference}.
Fig.\ref{LasingFreq}(a) presents lasing frequencies of slab WG TE and TM modes (blue and red circles, respectively) in suspended photonic liquid lattices for several values of the interference angle $\theta$, as well as lasing frequencies of SPP modes propagating within plasmonic liquid lattices formed by four interfering SPPs with a relative interference angle set by $\alpha$ (red squares). Interestingly, unlike the 1D case the lasing frequency is not a monotonic function of the relative angle between the interfering waves.
For simplicity, in our simulations the laser material is set as a linear gain Lorentz model \cite{oughstun2003lorentz}, described by a dielectric function given by $\epsilon (f) = \epsilon_{l}+\epsilon_{L} \omega_{0}^{2}/(\omega_{0}^{2} - 4 \pi i \delta_{0} f -(2 \pi f)^{2})$ with resonance frequency centered at $\omega_{0}/(2 \pi) = 374.97$ THz ($800$ nm), liquid permittivity $\epsilon_{l}= n_{l}^{2} = (1.409)^{2} = 1.9852$, Lorentz linewidth $\delta/(2 \pi) = 52.521$ THz, and Lorentz permittivity $\epsilon_{L} = -0.0075$; the metal is taken as gold.

\section*{Discussion}

In this work we theoretically investigated optical properties of configurable liquid-made lattices, formed by interfering SPPs or slab WG modes. We leveraged the underlying complex nonlocal and nonlinear interaction described by the Ginzburg-Landau and Schr\"{o}dinger equations, which capture the effect of the self-induced action of the optical mode on itself due to geometrical changes of the gas-liquid interface, to predict formation of optical liquid lattices and bandstructure tuning due to symmetry breaking of hexagonal symmetric and square symmetric lattices as well as phase transition effects between hexagonal symmetric to face centered symmetric lattices.
We then applied the bandstucture tuning to demonstrate control over various properties of the lasing systems such as gain threshold, lasing frequency and emission direction of the corresponding lasing mode. 
Notably, the self-induced lasing threshold of the TLD film interacting with the surface optical mode, is expected to admit much lower values as compared to a similar liquid system without gas-fluid interface and therefore has the prospect
to serve as a future \textit{Lab on a Film} bio-sensing platform which integrates liquid delivery with self-induced DFB lasing mechanism.
Interestingly, the formation of a graphene-like liquid lattice with tunable Dirac points in lattices with hexagonal broken symmetry, is substantially different from other configurable platforms introduced to date for the formation of Dirac points in optical lattices, such as schemes which incorporate cold atoms \cite{zhu2007simulation} and Fermi gases \cite{tarruell2012creating}.
Additionally, since metamaterials can be utilized as an optical computational platform, as was recently demonstrated in a solid-made and non-reconfigurable setup for solution of linear equations \cite{estakhri2019inverse}, the adaptive property of optical liquid lattices has the potential to allow reconfigurable computation of computationally challenging problems of systems of linear and nonlinear algebraic equations (e.g., Eq.(\ref{CGLEplaneWave})), and also has the potential to serve as an emulator of many-body quantum mechanical problems such as electron propagation in an atomic lattice, including topological edge-state effects.  
We hope that our work will stimulate future experimental and theoretical studies to realize optical liquid lattices and to explore underlying nonlinear light-liquid interaction mechanisms which include also birefringence and magneto-optical effects.

\textbf{Acknowledgments:} 

SR cordially thanks to Dr. Anna Rubin-Brusilovski for valuable and inspiring discussions.

This work was supported by the Defense Advanced Research Projects Agency (DARPA) DSO’s NAC (HR00112090009)  and NLM, the Office of Naval Research (ONR) Multidisciplinary University Research Initiative (MURI), the National Science Foundation (NSF) grants CCF-1640227, the Semiconductor Research Corporation (SRC), and the Cymer Corporation. 









\widetext
\begin{center}
\newpage
\title{SI}
\textbf{Supplementary Material for: \\ ``Nonlinear, tunable and active optical metasurface with liquid film''}

\text{Shimon Rubin and Yeshaiahu Fainman}

\textit{Department of Electrical and Computer Engineering, University of California, San Diego, 9500 Gilman Dr., La Jolla, California 92023, USA}
\end{center}

\setcounter{equation}{0}
\setcounter{figure}{0}
\setcounter{section}{0}
\setcounter{table}{0}
\setcounter{page}{1}

\renewcommand{\thesection}{S.\arabic{section}}
\renewcommand{\thesubsection}{\thesection.\arabic{subsection}}
\makeatletter 
\def\tagform@#1{\maketag@@@{(S\ignorespaces#1\unskip\@@italiccorr)}}
\makeatother
\makeatletter
\makeatletter \renewcommand{\fnum@figure}
{\figurename~S\thefigure}
\makeatother
\makeatletter \renewcommand{\fnum@table}
{\tablename~S\thetable}
\makeatother

\section{Thin liquid dielectric film equation under thermocapillary and solutocapillary effects}

\subsection*{Thin liquid dielectric film equation under thermocapillary and solutocapillary effects}

The Navier-Stokes equations for a noncompressible Newtonian fluid of viscosity $\mu$, mass density $\rho$, velocity field $u_{i}$ and stress tensor $\tau_{ij}$ are given by [38]
\begin{equation}
	\rho \left( \partial_{t}u_{i}+ u_{j} \partial_{j} u_{i} \right) = \partial_{j} \tau_{ij}; \quad i,j=x,y,z;,
\label{NavierStokes}    
\end{equation}
the stress balance equation for the free surface of the thin liquid film is given by
\begin{equation}
	\tau_{ij} n_{j}  = \sigma n_{i} \vec{\nabla} \cdot \hat{n} - \vec{\nabla}_{\parallel} \sigma,
\label{MatchingConditions}    
\end{equation}
where, $\sigma$ is the surface tension, $\vec{\nabla} \cdot \hat{n}$ is the divergence of the normal 
and $\vec{\nabla}_{\parallel}$ stands for a gradient with respect to the in-plane coordinates ($y,z$).
Applying low Reynolds number and thin film approximations (i.e., lubrication approximation), 
allows to neglect the inertial terms in the Navier-Stokes equations, Eq.(S\ref{NavierStokes}), and drop the in-plane derivatives
relative to the normal derivative. Together with the thin film limit
of the matching conditions, Eq.(S\ref{MatchingConditions}) [64], yields the following equation for the thin film deformation $\eta$,
as a function of spatial variations of the surface tension
\begin{equation}
	\dfrac{\partial \eta}{\partial t} + D_{\sigma} \nabla^{4}_{\parallel} \eta = -\dfrac{h_{0}^{2}}{2 \mu} \nabla^{2}_{\parallel} \sigma; \quad D_{\sigma} \equiv \sigma_{0}^{ } h_{0}^{3}/(3 \mu)
\label{ThinFilmST}	
\end{equation}
The case where $\sigma$ depends on temperature was presented in [36], and below we consider the case where $\sigma$ depends on concentration of molecular species. 

\textbf{Solutocapillary driven deformation of TLD film}

Assume a solution comprised of a photoactive molecule and passive solute ions. In particular assume that the photo-active specie can switch between two states labeled as $A$ and $B$ and that the surface tension can be written as a function of the two species concentrations $c_{A,B}$, as
\begin{equation}
	\sigma(c_{A},c_{B})=\sigma_{0}-\sigma_{c}^{(A)} (c_{A}^{ }-c_{A}^{0}) - \sigma_{c}^{(B)} (c_{B}^{ } -c_{B}^{0}).
\label{SurfaceTensionGradConc}
\end{equation}
Here, $c_{A,B}^{0}$ are the initial concentrations and $\sigma_{c}^{(A,B)}$ are the corresponding Marangoni coefficients due to concentration changes of the two species. Furthermore, assuming low species concentration, the dynamics of the ionic species is governed by the Nernst-Planck (mass conservation) equations 
\begin{equation}
\begin{split}
	\dfrac{\partial c_{A}}{\partial t} + \vec{\nabla} \cdot \vec{J}_{A} = k_{on}^{ } c_{B} - k_{off}^{ } c_{A}
\\
	\dfrac{\partial c_{B}}{\partial t} + \vec{\nabla} \cdot \vec{J}_{B} = k_{off}^{ } c_{A} - k_{on}^{ } c_{B},
\end{split}	
\label{NP}
\end{equation}
where the mass flux density currents $\vec{J}_{A,B}$, are given by $\vec{J}_{A,B}=-D_{A,B} \vec{\nabla} c_{A,B} + c_{A,B} \vec{u}_{A,B}$. Here, $D_{A,B}$ are the molecular diffusion coefficients, $\vec{u}_{A,B}$ are advection velocity fields, whereas $k_{on}$ and $k_{off}$ are the corresponding intensity and wavelength dependent coefficients that determine conversion rate from $B$ to $A$ and from $A$ to $B$, respectively.   
Taking half-sum and half-difference of Eq.(S\ref{NP}), and furthermore assuming for simplicity that both species admit identical molecular diffusion coefficients, i.e., $D_{A}=D_{B} \equiv D$, one can readily verify that a constant solution for the mean concentration, $c=c_{0}$, where $c$ is defined as $c \equiv (c_{A}+c_{B})/2$, satisfies the following homogeneous Nernst-Planck equation
\begin{equation}
	\dfrac{\partial c}{\partial t} + D \nabla^{2}_{\parallel} c = 0,
\end{equation}
whereas half concentration difference $\overline{c}$, defined as  $\overline{c} \equiv (c_{A}-c_{B})/2$, is governed by the following non-homogeneous Nernst-Planck equation
\begin{equation}
	\dfrac{\partial \overline{c}}{\partial t} - D \nabla^{2}_{\parallel} \overline{c}=2k_{on}^{ }(c_{0}-\overline{c})-2k_{off}^{ }(c_{0}+\overline{c}).
\label{NernstPlanckRho}    
\end{equation}
For the case where initial concentrations of $A$ and $B$ species are equal, the initial value of $\bar{c}$ is zero and small variations of this quantity describe conversion between the two molecular configurations.
Utilizing the relations $\sigma_{c}^{A}c_{A}^{ }+\sigma_{c}^{B}c_{B}^{ }=\sigma_{c} c + \bar{\sigma}_{c} \bar{c}$ 
where $\sigma_{c}^{ } \equiv (\sigma_{c}^{A}+\sigma_{c}^{B})/2$ is half-sum and $\bar{\sigma}_{c}^{ } \equiv (\sigma_{c}^{A}-\sigma_{c}^{B})/2$ is half-difference of the corresponding solutocapillary Marangoni coefficients, allows to express 
the surface tension $\sigma(c_{A},c_{B})$ given by Eq.(S\ref{SurfaceTensionGradConc}), as a function of concentration difference via
\begin{equation}
	\sigma(c_{A},c_{B}) = \sigma_{0} - \sigma_{c}^{A} (c_{A}^{ } - c_{A}^{0}) - \sigma_{c}^{B} (c_{B}^{ } - c_{B}^{0}) = \sigma_{0}  - \overline{\sigma}_{c} (\overline{c} - \overline{c}_{0}).
\label{sigmaAB} 
\end{equation}
Substituting the expression for the surface tension given by Eq.(S\ref{sigmaAB}) into Eq.(S\ref{ThinFilmST}) yields
\begin{equation}
	\dfrac{\partial \eta}{\partial t} +D_{\sigma} \nabla^{4}_{\parallel}  \eta = \dfrac{h_{0}^{2} \bar{\sigma}_{c}^{ }}{2 \mu} \nabla_{\parallel}^{2} \bar{c}.
\label{ThinFilmEqSC}	
\end{equation}
In a regime characterized by  
\begin{equation}
	\tau_{el} \ll \tau_{c} \ll \tau_{l},
\label{hierarchy}    
\end{equation}
where $\tau_{c}$ is the molecular diffusion time scale defined by $\tau_{c} \equiv d^{2}/D$, 
one can overlook the time derivative in Eq.(S\ref{NernstPlanckRho}) provided the observation time scale is on the order (or larger) than $\tau_{l}$. In particular, $\tau_{c} \ll \tau_{l}$ regime implies that the corresponding film thickness satisfies $h_{0}^{3} \ll 3 \mu D d^{2}/\sigma_{0}$.
Furthermore, assuming a negligible $k_{off}$, which corresponds to a case of negligible optically stimulated reaction $B \rightarrow A$, and utilizing Eq.(S\ref{NernstPlanckRho}) to eliminate the laplacian term in Eq.(S\ref{ThinFilmEqSC}) yields
\begin{equation}
	\dfrac{\partial \eta}{\partial t} + D_{\sigma} \nabla^{4}_{\parallel}  \eta = - \dfrac{h_{0}^{2} \overline{\sigma}_{c}^{ }}{\mu D} k_{on} (c_{0} - \bar{c}) \simeq - \text{Ma}_{c} \dfrac{h_{0}}{\tau_{on}} \dfrac{c_{0}}{\Delta c^{(0)}} \dfrac{I}{I_{0}}.
\label{ThinFilmSCsurftension}	
\end{equation}
Here, $\text{Ma}_{c} \equiv h_{0} \bar{\sigma}_{c} \Delta c^{(0)}/(\mu D)$ is the solutocapillary Marangoni number; $\tau_{on}$ is the characteristic time scale of the conversion process of $B$ to $A$, defined as $\tau_{on} \equiv 1/k_{on}$; $\kappa$ is the $k_{on}$ coefficient normalized by the optical intensity, defined via the relation $1/\tau_{on} \equiv k_{on}  \equiv \kappa I$; $k_{on}^{(0)} \equiv \kappa I_{0}$; $1/\tau_{on}^{(0)} \equiv \kappa I^{(0)} $. In the last equality in Eq.(S\ref{ThinFilmSCsurftension}), we have assumed that $\bar{c} \ll c_{0}$ and kept the leading term in the $\bar{c}/c_{0}$ expansion. Since Eq.(S\ref{ThinFilmSCsurftension}) is a linear differential equation with respect to $\eta$, we can represent the solution in terms of the corresponding Green's function given by Eq.(3), 
where for simplicity we have removed the superscript $(0)$ from $\Delta c^{(0)}$ and $\tau_{on}^{(0)}$. Note that in our derivation we assumed that $c_{A}$ and $c_{B}$ are uniform along the vertical direction; in practice ions distributions could be different as the affinity of each ionic species to the liquid-gas interface could be different. 

\textbf{Thermocapillary driven deformation of TLD film}

The heat transport 2D equation is given by
\begin{equation}
	\dfrac{\partial T}{\partial t} - D_{th}^{ } \nabla^{2}_{\parallel} T= \dfrac{\Delta T}{I_{0} \tau_{th}}  \chi  I; \quad \chi \equiv \dfrac{\alpha_{th}^{ }  d^{2} I_{0}}{k_{th}^{ } \Delta T},
\label{HeatEquationDiffusion2c}    
\end{equation}
where $D_{th}^{ }=k_{th}^{ }/(\rho^{ } c_{p}^{ })$ is the heat diffusion coefficient;  $\rho^{ }$, $c_{p}^{ }$, $k_{th}^{ }$, $\alpha_{th}^{ }$ are the mass density, specific heat,
heat conductance, and optical absorption coefficient, respectively; $\tau_{th}=d^{2}/D_{th}^{ }$ is the typical heat diffusion time scale; $d$ is the typical length scale along the in-plane direction; $\chi$ is the dimensionless intensity of the heat source, and $I_{0}$ is the characteristic optical intensity.
Furthermore, we assume that the surface tension depends on the temperature via 
\begin{equation}
	\sigma(T)=\sigma_{0}-\sigma_{T} \Delta T; \quad \Delta T \equiv T-T_{0},
\label{SurfaceTensionGrad}
\end{equation}
where $\sigma_{T}$ is the Marangoni constant, 
and $\sigma_{0}$ is the surface tension at temperature $T_{0}$.
Inserting the constitutive relation for the surface tension Eq.(S\ref{SurfaceTensionGrad}) into Eq.(22) in the main text,
yields the following equation for the thin film deformation,
\begin{equation}
	\dfrac{\partial \eta}{\partial t} + D_{\sigma} \nabla^{4}_{\parallel}  \eta=  \dfrac{\sigma_{T} h_{0}^{2}}{2 \mu}  \nabla^{2}_{\parallel} T,
\label{ThinFilmEqU33}
\end{equation} 
which includes the effects of surface tension and thermocapillarity (gravity typically admits a negligible effect on the dynamics of TLD film on small scales, and is expected to have a non-negligible effect on a much larger capillary length scale). 
Note that the small thickness of the fluid film implies that the temperature distribution along the vertical direction is mostly determined by the heat transport in the bulk rather than surface transport through the gas-fluid interface (see more details Supplementary Material [36]). 

The typical values of the time scales $\tau_{l}$, $\tau_{th}$ and $\tau_{el}$, that govern the transport of liquid, heat and propagation of SPP, respectively, are subject to the following hierarchy [36]
\begin{equation}
	\tau_{el} \ll \tau_{th} \ll \tau_{l} = d^{4}/ D_{\sigma}.
\label{hierarchy}    
\end{equation}
Considering "slow" processes of characteristic time scale $\tau_{l}$ allows to treat the dynamics of optical and heat transport (i.e. "fast" processes) as quasi-static, keep the time derivative in the thin film equation Eq.(S\ref{ThinFilmEqU33}) and eliminate the time derivative in Eq.(S\ref{HeatEquationDiffusion2c}). The latter allows to express the temperature laplacian in terms of optical intensity, which upon substitution into the left-hand side of Eq.(22) in the main text leads to [36]
\begin{equation}
	\dfrac{\partial \eta}{\partial t} + D_{\sigma} \nabla^{4}_{\parallel}  \eta=  - \dfrac{h_{0}}{2\tau_{th}} \text{Ma} \cdot \chi \cdot \dfrac{I}{I_{0}}.
\label{ThinFilmEqU}
\end{equation}

\section{Green's function for thin liquid dielectric film deformation}

The governing equations for TLD film deformation due to TC and SC mechanisms, are given by Eq.(S\ref{ThinFilmEqU}) and Eq.(30) in the main text, respectively. 
In a particular case of transient actuation, i.e., due to a source described by Heaviside step function $H(t)$, the corresponding governing equation is given by
\begin{equation}
	\dfrac{\partial \eta_{H}}{\partial t} - A \nabla^{2} \eta_{H} + B \nabla^{4} \eta_{H} = \delta(y-y_{0}) \delta(z-z_{0}) H(t),
\label{EtaH}	
\end{equation}
where $\eta_{H}= \int dt G_{l}$ is the indefinite time integral of the Green's function $G_{l}$, where $G_{l}$ is the Green's function of Eq.(S\ref{EtaH}) with $H(t)$ replaced by Dirac's delta function $\delta(t)$.
Here, $A= - \rho g h_{0}^{3}/(3 \mu), B= - \ell_{c}^{2} A$, and $\eta_{H}$ and $(y_{0},z_{0})$ is the actuation point.
The corresponding boundary conditions of the relevant rectangular domain $0 \leq y \leq d_{y}$, $0 \leq z \leq d_{z}$, are given by $\vec{\nabla} \eta_{H} \cdot \hat{n} = 0$ at $y=0,d_{y}$ and $z=0,d_{z}$ whereas the initial condition is $\eta_{H}(y,z,t=0) =0$. The corresponding deformation is then given by [36]
\begin{equation}
\begin{split}
	\eta_{H}(\vec{r}_{\parallel},t) = &  \sum\limits_{n,m=1}^{\infty} 
	\dfrac{d_{y} d_{z}}{2} \varphi_{m,n}(\vec{r}_{0\parallel}) f_{m,n}(t) \varphi_{m,n}(\vec{r}_{\parallel}) +
	\\
	& \sum\limits_{m=1}^{\infty} \sqrt{\dfrac{d_{y}}{2}}  \varphi_{m,0}(\vec{r}_{0\parallel})  f_{m,0}(t) \varphi_{m,0}(\vec{r}_{\parallel}) +
	\\
	& \sum\limits_{n=1}^{\infty} \sqrt{\dfrac{d_{z}}{2}}  \varphi_{0,n}(\vec{r}_{0\parallel})  f_{0,n}(t) \varphi_{0,n}(\vec{r}_{\parallel}).
\end{split}
\label{GreenEtaH}	
\end{equation}
Here, $f_{m,n}(t) \equiv (1-e^{- \lambda_{m,n} t})/\lambda_{m,n}$, 
\begin{equation}
	\lambda_{m,n} = \pi^{2} \left( A \left( \dfrac{m^{2}}{d_{y}^{2}} + \dfrac{n^{2}}{d_{z}^{2}} \right) + B \pi^{2} \left( \dfrac{m^{4}}{d_{y}^{4}} + 6 \dfrac{m^{2}}{d_{y}^{2}} \dfrac{n^{2}}{d_{z}^{2}}  +\dfrac{n^{4}}{d_{z}^{4}} \right) \right).
\end{equation}
and $\varphi_{m,n}(y,z)$ are the eigenfunctions of the corresponding Sturm-Liouville problem
\begin{equation}
\begin{split}
	\varphi_{m,n}(y,z) &= \dfrac{2}{d_{y} d_{z}} \cos \left( \dfrac{m \pi y}{d_{y}} \right) \cos \left( \dfrac{n \pi z}{d_{z}} \right)
\\	
	\varphi_{m,0}(y) &= \sqrt{\dfrac{2}{d_{y}}} \cos \left( \dfrac{m \pi y}{d_{y}} \right) 
\\
	\varphi_{0,n}(z) &= \sqrt{\dfrac{2}{d_{z}}} \cos \left( \dfrac{n \pi z}{d_{z}} \right). 
\end{split}	
\end{equation}
In our work we consider thin film response which evolve over time scales larger than $1/\lambda_{m,n}$ which is equivalent to taking $f(t \rightarrow \infty) = 1$.

\section{Derivations of Eq.6, Eq.7 and Eq.10}

Consider an interference of $N$ plane waves of real-valued amplitudes $a_{n}$, which propagate in the $y-z$ plane along the directions formed by angles $\theta_{n}$ which are measured relative to the positive direction of the $y$-axis. The associated wavevectors are given by $\vec{k}_{n}=k_{0}\left(\cos(\theta_{n}), \sin(\theta_{n}) \right)$, where $k_{0}=2 \pi /\lambda$ and $\lambda$ is the wavelength in the corresponding media (reduced by the effective refractive index factor), and the corresponding optical intensity $I$ is given by
\begin{equation}
	I = \Big \vert \sum_{n=0}^{N-1} a_{n} e^{i  \vec{k}_{n} \cdot \vec{r}_{\parallel} }  \Big \vert^{2} = I_{N} + 2\sum			\limits_{n \neq m}^{N-1} a_{n}a_{m} \cos \left( (\vec{k}_{n}-\vec{k}_{m}) \cdot \vec{r}_{\parallel} \right); \quad I_{N} 		\equiv \sum_{n=0}^{N-1} \vert a_{n} \vert^{2}.
\label{Intensity}	
\end{equation}
Consider the simplest $N=2$ case, that leads to a 1D optical liquid lattice. The intensity distribution of two optical plane waves  of equal intensity $I_{0}$ propagating along the $y-z$ plane with wave vectors $\vec{k}_{\pm} = k (\cos (\theta), \sin(\theta))$ (with relative angle $\Delta \theta = \pi - 2 \theta$), $\sqrt{I_{0}} e^{i k (\pm \cos(\theta) x +\sin(\theta) y)}$, is given by $2I_{0}(1+\cos(K(\theta) x))$, where $K(\theta) \equiv 2 k \cos(\theta)$ is the corresponding wave vector.
Inserting the resultant intensity into Eq.(2) in the main text and utilizing the relation $\int dt G_{l} = \eta_{H}$, where $\eta_{H}$ is given by [36] (see also Supplementary Information), singles out the $\bar{n}$ term in the series expansion of $\eta_{H}$, where $\bar{n} = 2d/\lambda $ is an integer, and yields the corresponding 1D deformation of the TLD film (for $t \rightarrow \infty$)
\begin{equation}
	\dfrac{\eta(x)}{h_{0}}  = - \dfrac{2 \text{M}}{\tau_{th} \lambda_{\bar{n}}} \cos(K(\theta) x); \quad K(\theta) = \dfrac{2 \pi}{\Lambda(\theta)}; \quad \Lambda (\theta) = \dfrac{\pi}{k \cos(\theta)}.
\label{PeriodicDeformation}
\end{equation}

For the case $N=3$ where the relative amplitudes of the three beams satisfy
\begin{equation}
	a_{1}=a_{2}=a_{3}/q \equiv a,
\label{RatioQ}	
\end{equation}
where $a$ and $q$ are real constants, 
and furthermore assuming the corresponding propagation directions are given by $\hat{n}_{0}= \cos(\theta) \hat{y} - \sin(\theta) \hat{z}$, $\hat{n}_{1}= -\cos(\theta) \hat{y} - \sin(\theta) \hat{z}$, $\hat{n}_{2}=\hat{z}$ (which can be also written as $\theta_{n} =(n+1) \pi + (-1)^{n} \theta $ (n=0,1), $\theta_{2}=\pi/2$), yields the following intensity distribution
\begin{equation}
	I/a^{2} = 1 + 2q^{2} + 2q^{2} \cos \Big[ \dfrac{4 \pi y}{\Delta y} \Big] +4 q \cos \Big[ \dfrac{2 \pi y}{\Delta y} \Big] 
	\cos \Big[ \dfrac{2 \pi z}{\Delta z} \Big].
\label{HexIntensity}	
\end{equation}
Here, $\Delta y$ and $\Delta z$ are the corresponding periodicities along the $y$ and $z$ directions, respectively, given by
\begin{equation}
	\Delta y = \dfrac{\lambda}{\cos(\theta)}; \quad
	\Delta z = \dfrac{\lambda}{1+\sin(\theta)},
\label{HexIntensityPer}	
\end{equation}
which depend on the angle $\theta$ and the wavelength $\lambda$, but do not depend on $a$ nor on $q$.
Inserting the intensity given by Eq.(S\ref{HexIntensity}) into Eq.(1) in the main text, yields the following TLD film TC-driven deformation (for $t \rightarrow \infty)$
\begin{equation}
\begin{split}
 	\dfrac{\eta(\vec{r}_{\parallel})}{h_{0}} =- \dfrac{\eta_{max}}{3} \Big[ \cos \left( \dfrac{4 \pi y}{\Delta y} \right) + 	
 	2\cos \left( \dfrac{2 \pi y}{\Delta y} \right) \cos \left( \dfrac{2 \pi z}{\Delta z} \right) 
 	 \Big], 
\end{split}
\label{HexDeformation}	  
\end{equation}
where $\eta_{max} \equiv (3\bar{q} \text{M})/(\lambda_{\bar{n},0} \tau_{th})$, $\bar{m},\bar{n}$ are integers that satisfy $\bar{m}=2 d_{x}/\Delta x$, $\bar{n}=2 d_{y}/\Delta y$, and $\bar{q}=2\lambda_{2\bar{n},0}/\lambda_{\bar{m},\bar{n}}$. 



An optical liquid lattice with a rectangular tunable symmetry can be formed by interfering four beams where the relative amplitudes satisfy 
\begin{equation}
	a_{0,2}=q a, a_{1,3}=a,
\end{equation}
and the propagation angles are given by $\theta_{n}=(1+2n) \pi/4+(-1)^{n}\alpha$ ($n=0,1,2,3$), where $\alpha$ is a real parameter. Under these assumptions the optical intensity, Eq.(S\ref{Intensity}), is given by 
 \begin{equation}
 \begin{split}
 	 I/(4a^{2}) =  \cos \left( \dfrac{\pi y}{\Delta y} - \dfrac{\pi z}{\Delta z}  \right)^{2} +	
 			 q \left( \cos \left( \dfrac{2\pi y}{\Delta y} \right) + \cos \left( \dfrac{2 \pi z}{\Delta z} \right) \right)+ q^{2} \cos \left( \dfrac{\pi y}{\Delta y} + \dfrac{\pi z}{\Delta z}  \right)^{2}
\end{split}
\label{IntensitySquare}
\end{equation}
 where the periodicities along the $y,z$ directions are given by
\begin{equation}
 	\Delta y = \dfrac{\lambda}{2 \left( \cos(\alpha) - \sin(\alpha) \right)}; \quad \Delta z = \dfrac{\lambda}{2 \left( \cos(\alpha) + \sin(\alpha) \right)}.
 \label{SquarePeriod}	
\end{equation}
Inserting the intensity given by Eq.(S\ref{IntensitySquare}) into Eq.(3) in the main text, and following similar steps as described for the case $N=3$ above, yields the following deformation of the TLD film
\begin{equation}
 \begin{split}
 	\dfrac{\eta(\vec{r}_{\parallel}}{h_{0}} = - \dfrac{\eta_{max}}{3} \Big[ \cos \left( \dfrac{2 \pi y}{\Delta y} \right) + 
 	\cos \left(\dfrac{2 \pi z}{\Delta z}\right) +	
 	\cos \left( \dfrac{2 \pi y}{\Delta y} \right) \cos \left( \dfrac{2 \pi z}{\Delta z} \right) 
 	  \Big] 
 \end{split}
 \label{SquareDeformation}	  
\end{equation}
 where $\eta_{max} \equiv 3\text{M}(\bar{q}-3/4)/(2\lambda_{\bar{n},0} \tau_{th})$, $\bar{m},\bar{n}$ are integers that satisfy $\bar{m}=2 d_{y}/\Delta y$ and $\bar{n}=2 d_{z}/\Delta z$ and
 $\bar{q}$ satisfies the quadratic equation $(1/2-q^{2}/4)/(2\lambda_{\bar{m},\bar{n}})=(q-3/4)/\lambda_{\bar{n},0}$.
 
\section{Phase transition between 2D hexagonal to 2D face centered fluid lattice}

\begin{figure*}
	\includegraphics[width=\textwidth]{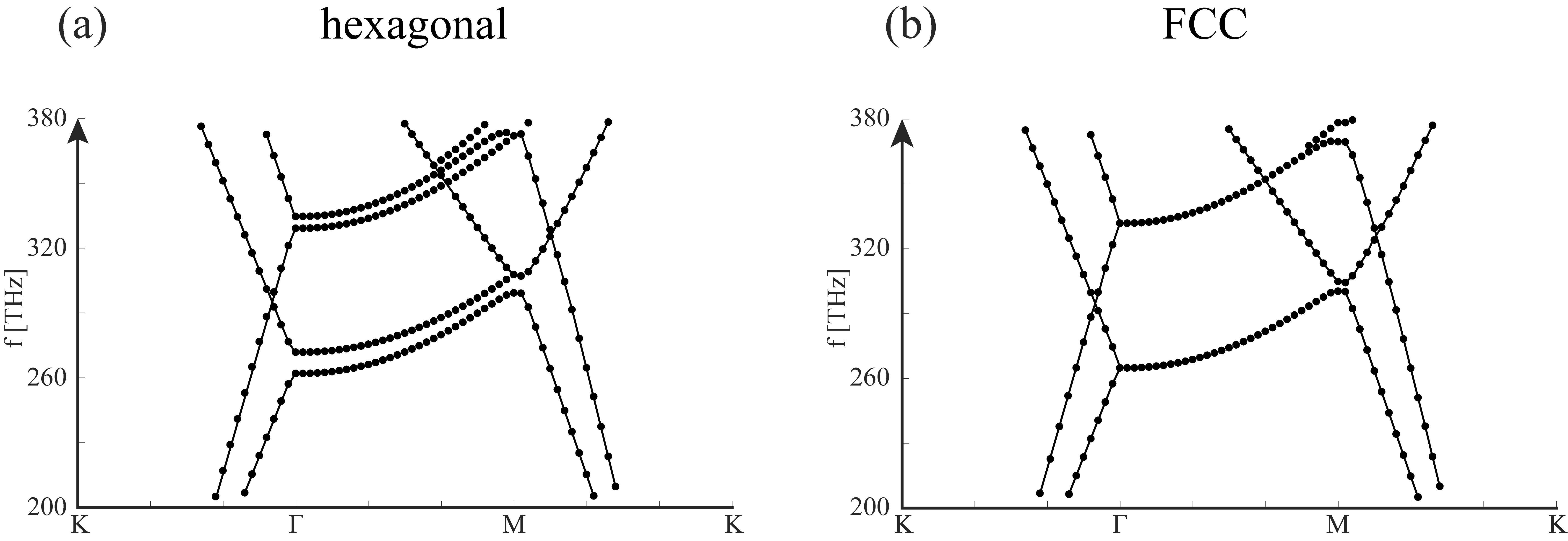}
    \caption{Numerical results presenting the effect of symmetry change of suspended photonic fluidic crystal on the corresponding TE waveguide modes: (a) hexagoonal symmetry (b) face-centered symmetry.}
    \label{HexagonalFCC}
\end{figure*}

Eq.(S\ref{HexIntensity}) describes intensity pattern due to interfering three plane waves where the parameter $q$ is defined by Eq.(S\ref{RatioQ})
Initially, the liquid lattice is formed by three interfering plane waves of equal amplitude and admits a hexagonal symmetry symmetry described in Fig.2a in the main text. By reducing the amplitude of the beam $n=2$, i.e., by taking increasingly smaller values of the parameter $q$, one is led towards merging of the liquid peaks (Fig.2h), and eventually to a face centered 2D optical liquid lattice presented in Fig.2i in the main text. Fig.S\ref{HexagonalFCC}a and Fig.S\ref{HexagonalFCC}b describe the initial bandstructure corresponding to optical liquid lattice of hexagonal symmetry and the terminal bandstructure of optical liquid with face centered symmetry. 

In the numerical simulation the dielectric properties are: refractive index $1.409$, $\eta_{max}=360$ nm $\eta_{min}=150$ nm, i.e, the maximal thickness of the suspended photonic liquid lattice is $720$ nm whereas the minimal thickness is $300$ nm.

\section{Depth averaged index for SPP and WG modes}

\textbf{SPP mode:}

The depth averaged index weighted by an exponential factor $2 q_{d} e^{-2 q_{d} x}$ is given by [36]
\begin{equation}
 		n_{D}(\eta(\vec{r}_{\parallel})) = 2 q_{d} \int\limits_{0}^{\infty} n(x) e^{-2 q_{d} x} dx = n_{l} + (n_{g} - n_{l}) e^{- 2 q_{d} (h_{0} + \eta(\vec{r}_{\parallel})}
\end{equation}
where $n_{l}$ and $n_{g}$ are the indices in the regions $0 < x < h$ and $x > h$, respectively.
Assuming the deformation is small relative to the the characteristic decay length of the SPP into the bulk, i.e. $q_{d} \eta \ll 1$, and keeping 
up to the linear term in $q_{d} \eta(\vec{r}_{\parallel})$, yields 
\begin{equation}
	n_{D}(\eta(\vec{r}_{\parallel})) = n_{0} + \Delta n_{D}; \quad \Delta n_{D} \equiv b \eta(\vec{r}_{\parallel})/h0
\label{FirstOrderChangeIndex}	
\end{equation}
where $n_{0}$ and $b$ are coefficients given by
\begin{equation}
	n_{0} = n_{l} + (n_{g} - n_{l})e^{-2 q_{d} h_{0}}; \quad b= 2 q_{d} h_{0} \big[ (n_{l} - n_{g}) - 1 \big].
\end{equation}	

\textbf{WG mode:} 

The depth averaged index, $n_{D}(\eta(\vec{r}_{\parallel}))$ weighted by intensity distribution of TE fundamental mode (see [56] for the corresponding field profile in slab WG configuration), is given by
\begin{equation}
	n_{D}(\eta(\vec{r}_{\parallel})) = 2/d \int\limits_{0}^{d/2} n_{l} \cos(h_{l}x)^{2} dx + \gamma_{g} \int\limits_{d/2}^{\infty} n_{g} \cos^{2}(h_{l}d/2) e^{2 \gamma_{g} (d/2 -x)} dx.
\end{equation}
The corresponding coefficients $n_{0}$ and $b$ which enter the expression Eq.(S\ref{FirstOrderChangeIndex}) for this case of a leading order change of the depth averaged index of TE slab WG mode is given
\begin{equation}
	n_{0} = n_{l} \big[ 1 + \text{sinc} (d_{0} h) \big] /2 + n_{g} \big[ 1 + \cos(d_{0} h) \big]/4; \quad b = \big[ \cos(d_{0}h)-\text{sinc}(d_{0} h) \big] n_{l}/2.
\end{equation}
where $\text{sinc}(d_{0} h) \equiv \sin(d_{0} h)/(d_{0} h)$. Similarly, analogous expressions can be obtained for fundamental TM mode.

\subsection*{Derivation of the complex nonlocal and nonlinear Ginzburg-Landau equation}

Eliminating the magnetic field from the following Maxwell equations (in SI units), $\vec{\nabla} \times \vec{E} = -\partial \vec{B}/ \partial t$ and $\vec{\nabla} \times \vec{H} = \partial \vec{D} / \partial t$, and furthermore assuming harmonic time dependence $e^{i \omega t}$ for all fields, yields
\begin{equation}
	\vec{\nabla} \times \left( \vec{\nabla} \times \vec{E} \right) = \vec{\nabla} \left( \vec{\nabla} \cdot \vec{E} \right) - \nabla^{2} \vec{E} = \epsilon_{r} \dfrac{\omega^{2}}{c^{2}} \vec{E}.
\label{Maxwell}
\end{equation}
Here, we have used the constitutive relations $\vec{D} = \epsilon \vec{E}$, $\vec{B} = \mu \vec{H}$ as well as $c^{2}=1/(\epsilon_{0} \mu)$, where $\epsilon \equiv \epsilon_{0} \epsilon_{r}$ and $\epsilon_{r}$ can depend on both frequency and is a functional of the electrical field $\epsilon_{r} = \epsilon_{r} \big[ \vert E \vert^{2} \big]$. The latter can be written as $\epsilon_{r} = \epsilon^{\prime}_{0} + i \epsilon^{\prime \prime}_{0} + \Delta \epsilon_{D} [A(\vec{r}_{\parallel}) ]$, where $\epsilon^{\prime}_{0}$ and $\epsilon^{\prime \prime}_{0}$ are the real and imaginary parts, respectively, of the depth averaged dielectric constant in a presence of undeformed TLD film, whereas $\Delta \epsilon_{D} [ A(\vec{r}_{\parallel}) ]$
is the convolution integral which captures the nonlocal and nonlinear response of the TLD film deformation, given by
\begin{equation}
	\Delta \epsilon_{d} [A(\vec{r}_{\parallel}) ]  = \chi \int d\vec{r}_{\parallel}^{\prime} G_{l}(\vec{r}_{\parallel}^{ },\vec{r}_{\parallel}^{\prime}) \vert A^{ } (\vec{r}_{\parallel}^{\prime}) \vert^{2},
\end{equation}
where $\chi$ is either $\chi_{TC}$ or $\chi_{SC}$ (see Eq.(4)).

The $x$ and $z$ components of Eq.(S\ref{Maxwell}) are given by
\begin{equation}
\begin{split}
	\dfrac{\partial^{2}E_{x}}{\partial y^{2}} + \dfrac{\partial^{2} E_{x}}{\partial z^{2}} - \dfrac{\partial^{2}E_{z}}{\partial x \partial z} + \epsilon_{r} E_{x} = 0
\\
	\dfrac{\partial^{2}E_{z}}{\partial y^{2}} + \dfrac{\partial^{2} E_{z}}{\partial x^{2}} - \dfrac{\partial^{2}E_{x}}{\partial x \partial z } + \epsilon_{r} E_{z} = 0.
\label{MaxXZ}
\end{split}
\end{equation}
Assuming a solution of the type
\begin{equation}
	\vec{E} = \big[ A(y,z) \vec{E}^{0}(x) + \delta \vec{E} \big] e^{i \beta z},
\end{equation}
where $\vec{E} = (E_{x}^{0},0,E_{z}^{0})$ is the electrical field of the unperturbed TM WG mode, and substitute it into Eq.(S\ref{MaxXZ}) we obtain
\begin{equation}
	L^{0} \delta \vec{E} = - L \vec{E}^{0}, 
\label{PerturbL}	 
\end{equation}
where $L_{0}$ and $L$ are given, respectively, by
\begin{gather}
L^{0}=
\begin{bmatrix} - \beta^{2} + \epsilon^{\prime}_{0}
& 
- i \beta \dfrac{\partial}{\partial x}
\\ 
- i \beta \dfrac{\partial}{\partial x}
&  
\dfrac{\partial^{2}}{\partial x^{2}}  + \epsilon^{\prime}_{0}
\end{bmatrix}
,
\end{gather}
and
\begin{gather}
L = 
\begin{bmatrix} \dfrac{\partial^{2}A}{\partial y^{2}} + 2 i \beta \dfrac{\partial A}{\partial z} + (i \epsilon^{\prime \prime}_{0} +\Delta \epsilon_{D} [A(\vec{r}_{\parallel}) ]) A 
& 
- i \beta A \dfrac{\partial}{\partial x}
\\ 
- i \beta A \dfrac{\partial}{\partial x}
&  
\dfrac{\partial^{2}A}{\partial y^{2}}  + (i \epsilon^{\prime \prime}_{0} + \Delta \epsilon_{D} [A(\vec{r}_{\parallel}) ]) A
\end{bmatrix}
.
\label{L}
\end{gather}
Here, $\vec{E}^{0}$ satisfies $L^{0}\vec{E}^{0} = 0$, and we have assumed the change of the mode is small on the plasmon wavelength $ \vert \partial^{2} A/ \partial z^{2} \vert \ll \vert 2 i \beta \partial A / \partial z \vert $, and that $\partial \delta \vec{E} / \partial y =\partial \delta \vec{E} / \partial z =0$.  

We employ the Fredholm theorem of alternative [65], which states that a non-zero solution $\delta \vec{E}$ to Eq.(S\ref{PerturbL}) exists only if the solution of the homogeneous system (i.e. solution of $L^{0} \delta \vec{E} = 0$) is orthogonal to the non-homogeneous term. Specifically, since $\vec{E}$ is the solution of homogeneous system (i.e. satisfies $L^{0}\vec{E}^{0}=0$), the condition above is given by
\begin{equation}
	\int\limits_{-\infty}^{\infty} dx \vec{E}^{0 \dagger} L \vec{E}^{0} = 0,
\label{Fredholm}	
\end{equation}
where $\vec{E}^{0 \dagger}$ stands for the complex conjugate of the transpose vector (see also [49] for application of Fredholm theorem of alternative for derivation of the governing equation of SPPs which propagate in tapered plasmonic waveguides). Inserting the expression for $L$ into Eq.(S\ref{Fredholm}) leads to the following expression under the integral
\begin{equation}
	\left( (E_{x}^{0})^{2} + (E_{z}^{0})^{2} \right) \dfrac{\partial^{2}A}{\partial y^{2}} + 2 i \beta (E_{x}^{0})^{2} \dfrac{\partial A}{\partial z} + (i \epsilon^{\prime \prime}_{0} + \Delta \epsilon_{D} [A(\vec{r}_{\parallel}) ])A \vert \vec{E}_{0} \vert^{4} - i \beta A \dfrac{\partial}{\partial x} \left( E_{x}^{0} E_{z}^{0} \right) = 0.
\label{BefDepthAvg}	
\end{equation}
Applying depth averaging, $\langle ... \rangle \equiv \int\limits_{-\infty}^{\infty}...dx/ \int\limits_{-\infty}^{\infty}dx$, to Eq.(S\ref{BefDepthAvg}), yields
\begin{equation}
	I \dfrac{\partial^{2}A}{\partial y^{2}} + 2 i \beta I_{x} \dfrac{\partial A}{\partial z} + i \Gamma A + N \Delta \epsilon_{D} [A] A = 0,
\end{equation}
where $I \equiv \langle \vert \vec{E}^{0} \vert^{2} \rangle$, $I_{x} \equiv \langle  \left( {E}_{x}^{0} \right)^{2} \rangle$, $\Gamma \equiv \langle \epsilon^{\prime \prime}_{0} \vert \vec{E}^{0} \vert^{4} \rangle$, and $N \equiv \langle \vert \vec{E}^{0} \vert^{4} \rangle$, and where the last term in Eq.(S\ref{BefDepthAvg}) vanishes after the integration due to vanishing of the fields at infinity.

\section{Bandstructure in hexagonal and rectangular plasmonic liquid lattices}

\begin{figure*}
	\includegraphics[width=\textwidth]{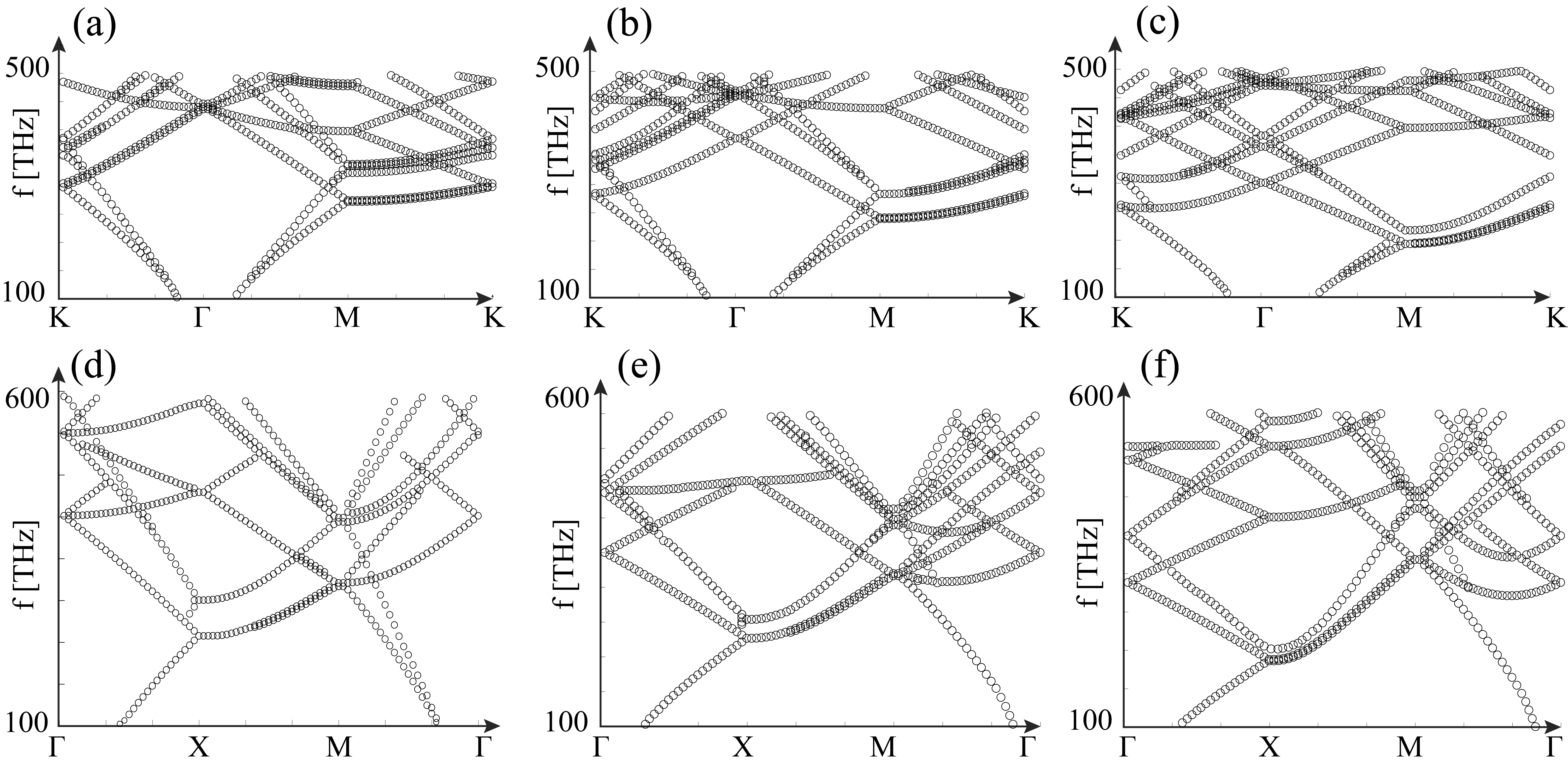}
    \caption{Numerical simulation results presenting the bandstructure of SPP modes in plasmonic liquid lattice with tunable hexagonal and rectangular symmetries formed by interfering SPPs. (a-c) Hexagonal symmetric liquid lattices that correspond to configurations $\theta=\pi/6$, $\theta=\pi/4$ and $\theta=\pi/3$, respectively. (d-f) Rectangular symmetric liquid lattices that correspond to the configurations (d) $\alpha=0^{o}$ (e) $\alpha=10^{o}$ and (f) $\alpha=20^{o}$. The corresponding deformation of the gas-fluid interface in all configurations is described in Fig.2.}
    \label{HexagonalRectangularSPP}
\end{figure*}

Fig.\ref{HexagonalRectangularSPP} presents numerical simulation results of the bandstructure formed by three and four interfering SPPs deforming a TLD film adjacent to a metal substrate (see Fig.1(a)), leading to a formation of a plasmonic liquid lattice with hexagonal and rectangular symmetry, respectively. Fig.\ref{HexagonalRectangularSPP}(a-c) present bandstructure modification due to change of the relative angle between the interfering beams; tuning $\theta$ transforms the hexagonally symmetric liquid lattice (Fig.\ref{HexagonalRectangularSPP}(a)) characterized by $\theta=\pi/6$, into liquid lattices of a broken hexagonal symmetry, characterized by $\theta=\pi/4$ (Fig.\ref{HexagonalRectangularSPP}(b)) and $\theta=\pi/3$ (Fig.\ref{HexagonalRectangularSPP}(c)). 
Fig.\ref{HexagonalRectangularSPP}(d-e) present the change of the SPP's bandstructure, due to corresponding modification of the underlying liquid lattice symmetry, schematically presented in Fig.2 in the main text; square symmetric lattice ($\alpha=0^{o}$) transforms to rectangular symmetric lattices ($\alpha=10^{o},20^{o}$).
  
\section{Self-induced index modulation and gain modulation in 1D optical liquid lattices}

Consider the 1D case of two counter propagating SPPs. In this case the periodic thin film deformation admits a sinusodial form, which introduces
changes ot the dielectric function of the type
\begin{equation}
	\Delta \epsilon = B \cos \left( \beta_{0}z/\sqrt{2} \right); \quad B= - \dfrac{\tau_{l} b }{16 \tau_{th} \lambda_{N}} \text{M}.
\end{equation}
Treating SPP excitation as a guided mode, the corresponding coupling coefficient between the two SPPs can be written as [55,56]
\begin{equation}
	\vec{E} = \sum_{m} A_{m}(z) \vec{E}_{m}(x,y)e^{-i(\omega t - \tilde{\beta}_{m}z)}.
\end{equation}
Assuming low dissipation leads to a set of coupled-mode equations which describe the variation in the mode amplitudes
under propagation along the $z$ direction in the waveguide. 
\begin{equation}
	 \pm \dfrac{da_{m}(z)}{dz} = i \sum\limits_{m,n} \kappa_{m n} a_{n}(z) e^{i(\tilde{\beta}_{m}-\tilde{\beta}_{n})},
\end{equation}
[52] where the positive and negative signs correspond to $\beta_{m}>0$ and $\beta_{m}<0$, respectively.
Here, $\kappa_{mn}$ is the coupling coefficient given by 
\begin{equation}
	\kappa_{mn} = \dfrac{\omega}{4}\int  \int \vec{E}^{*}_{m} \cdot \Delta \epsilon \cdot \vec{E}^{ }_{n} dx dz,
\end{equation}
which reflects an additional assumption that the longitudinal components of the guided modes are much smaller that the transversal components.
Employing fourier series expansion of the dielectric perturbation of period $\Lambda$
\begin{equation}
	\Delta \epsilon =e^{i \theta} \epsilon_{0} \sum\limits_{q} \epsilon_{q} e^{i q K z}; \quad K \equiv  2 \pi / \Lambda; \quad \theta \equiv \tan^{-1}(\epsilon^{\prime \prime}/\epsilon^{\prime}),
\end{equation}
and resonant approximation, i.e. neglecting the oscillation components, yields
the following coupled-mode equations describing the interaction between the forward and backward propagating waves
\begin{equation}
\begin{split}
	\dfrac{d}{dz}A_{+} &= i \kappa e^{i \theta} A_{-} e^{-2i \tilde{\Delta z}}
\\
	-\dfrac{d}{dz}A_{-} &= i \kappa e^{i \theta} A_{+} e^{2i \tilde{\Delta z}}.
\end{split}	
\end{equation}
Here,
\begin{equation}
	2 \tilde{\Delta} = 2 \Delta - i g; \quad 2 \Delta = 2 \beta - q K,
\end{equation}
$g>0$ is the gain coefficient and $\kappa$ is a real constant defined by
\begin{equation}
	 \kappa =\dfrac{ e^{i \theta}  \omega \epsilon_{0}}{4} \int \int \Delta \epsilon_{q} \vert \vec{E} \vert^{2} dx dy.
\end{equation}
The coupling coefficient $\kappa$ calculated for sinusoidal undulation, induced by the fundamental TE polarized modes, is given by [56]
\begin{equation}
	\kappa =k \eta_{M}; \quad k=e^{i \theta} \dfrac{k_{l}^{2}-\beta^{2}}{8 \beta (d+2/\gamma_{g})},
\end{equation}
where $\eta_{M}$ is the maximal deformation relative to the initial planar surface (i.e. half of the peak to peak depth), $\gamma_{g}^{2} = \beta^{2} - k_{g}^{2}$, and $k_{l,g}^{2} = \mu_{0} \epsilon_{l,g} \omega^{2} = n_{l,g}^{2} \omega^{2} / c^{2}$. 
For instance, values $d=0.85$ $\mu$m, $\lambda=0.85$ $\mu$m, $n_{l}=2.5$ which yield propagation constant of fundamental TE mode $\beta = 18.52$ $\mu$m$^{-1}$ (see Supplementary Information), and $\eta_{M}=182.5$ $\mu$m, the coupling coefficient between counter-propagating fundamental TE modes is $\kappa=-1.73 e^{i \theta}$.

Following the coupled mode theory [52,55], the corresponding reflection coefficient can be written as 
\begin{equation}
	\vert B(z)/A(0) \vert^{2} = F \sinh^{2}(\alpha(z-L))/(1+F \sinh^{2}(\alpha L)),
\end{equation}
where $F=1/(1-(\Delta/\kappa)^{2})$, $\alpha=\sqrt{\kappa^{2}-\Delta^{2}}$ and $\kappa$ is a coupling coefficient which is a linear function of the deformation, provided it is small relative to the slab WG thickness [52,55]. The case of maximal power transfer between the right and the left propagating modes, is characterized by $\Delta = 0$ which implies $F=1$ and $\alpha = \kappa$. Importantly, in our case the coupling coefficient is a function of intensity, and higher intensities lead to a higher temperature gradient and therefore to a higher deformation. Specifically, the coupling coefficient can be written as $\kappa = \kappa_{0} I_{0}$,
where $\kappa_{0}$ is the constant of proportionality between the coupling constant and field intensity, given by
\begin{equation}
	\kappa_{0} = k \overline{\text{M}} h_{0}; \quad \overline{\text{M}} =\dfrac{2 \text{Ma}}{\tau_{th} \lambda_{N}} \cdot \dfrac{\alpha_{th}^{ }  d^{2}}{k_{th}^{ } \Delta T}.
\end{equation}
Here, $h_{0}$ is the mean thickness of the TLD film, 
and $k$ depends on the specific modes of interest; e.g., for WG TE-TE coupling, the expression for $k$ is given in [56]. In the center of the fluidic slot $z=L/2$, the reflection then takes the form
\begin{equation}
	\vert B(z) \vert^{2} / I_{0} = \sinh^{2}(\kappa_{0} I_{0}L/2)/(1+\sinh^{2}(\kappa_{0} I_{0} L)).
\label{NonLinearDFB}	
\end{equation}
\begin{figure*}
	\includegraphics[width=\textwidth]{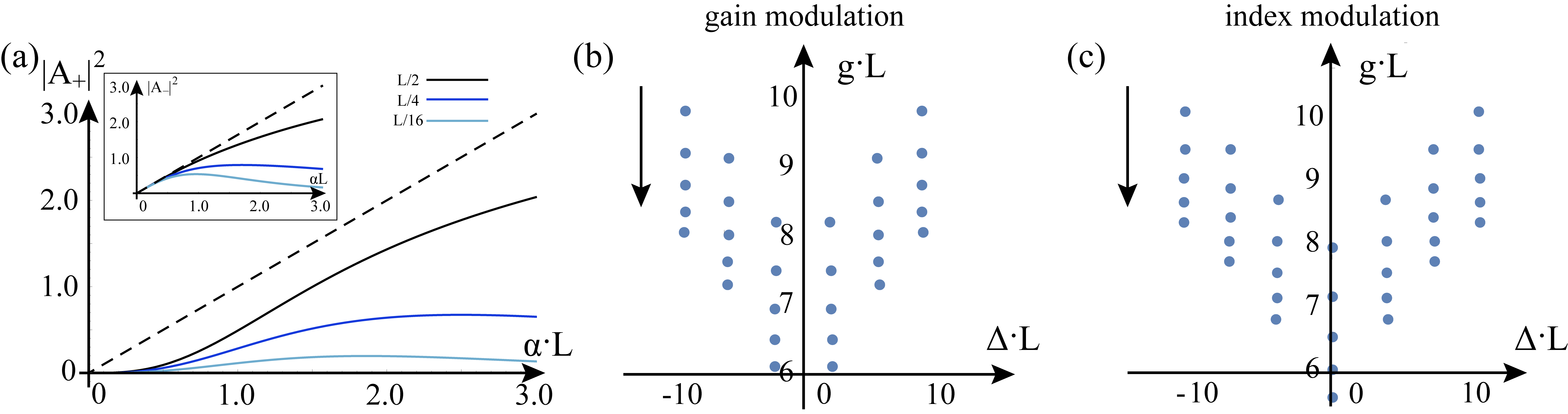}
    \caption{Numerical results presenting the effect of self-induced and nonlinear interaction between the right- and left-propagating modes of wave numbers $k_{\pm}$, leading to a 1D periodic deformation of a TLD film. Intensity dependent tuning of the amplitude leads to modulation of the corresponding coupling coefficient, transmission, reflection and lasing threshold. 
     (a) Non-linear intensity of the right- and left-propagating modes as a function of the optical intensity, described by Eq.(S\ref{NonLinearDFB}), along three different locations within the fluid cell. (b,c) A solution of Eq.(S\ref{DFBoscillationcond}) presenting the corresponding threshold as a function of a phase mismatch and as a function of the pump intensity leading to five different coupling constants. The arrow indicates the direction of increasing optical power, leading to a higher coupling constant and a lower threshold.}
    \label{DFBreflect}
\end{figure*}
Fig.\ref{DFBreflect}(a) presents the leading contribution of the self-induced TLD film periodic deformation on the reflection (see Eq.(S\ref{NonLinearDFB})) as a function of dimensionless intensity $\kappa_{0} I_{0} L$, presenting enhanced reflection of the mode $A_{+}$ for three different positions along the fluidic slot of total length $L$; the inset presents the corresponding intensity of the mode $A_{-}$. 
Changes of the coupling coefficient affect the oscillation condition for both index coupling and gain coupling modalities.  
The oscillation condition in DFB laser cavities for index coupling mechanism, can be written as the following complex equation [55] 
\begin{equation}
	\kappa L \sinh(i \gamma L) = \pm \gamma L; \quad  \gamma = -i \Big[ \left(\dfrac{g}{2} + i \Delta \right)^{2} + \kappa^{2} \Big]^{1/2},
\label{DFBoscillationcond}	 
\end{equation}
where the corresponding solution for $g$ and $\Delta$ must satisfy simultaneously vanishing of the real and the imaginary parts of Eq.(S\ref{DFBoscillationcond}). For the case of a gain coupling mechanism, a similar equation with $\kappa \rightarrow i \kappa$ holds. 
Fig.\ref{DFBreflect}(c,d) present the threshold as a function of the mismatch parameter $\Delta$, for a few cases of increasingly higher power which lead to corresponding higher values of the coupling coefficient (the direction of increasing power is marked by an arrow). While Fig.\ref{DFBreflect}(c) corresponds to an index coupled feedback mechanism, which is relevant for TLD films on top of flat gain films (e.g., Fig.2(f)), Fig.\ref{DFBreflect}(d) corresponds to a gain coupled feedback which can be realized for SPPs or WG modes that host a TLD film with gain (e.g. Fig.2(d,e)).    

\section{Temperature field and deformation of 1D plasmonic liquid lattice}

Consider the setup schematically described in Fig.1(d) where the width of the metal substrate coincides with the width of liquid slot $d_{z}$.
Under a pair of counter propagating SPP beams which generate intensity $I_{0} \cos^{2}(k(x-d_{z}/2))$,
the corresponding temperature field increase $\Delta T_{\infty}$ at times $t \ll \tau_{th} \equiv d_{z}^{2}/D_{th}$ is given by
\begin{equation}
	\Delta T_{\infty} = \dfrac{D_{th} \alpha_{th} I_{0}}{k_{th}} \int\limits_{0}^{d_{z}} dx_{0} T_{H}(x,x_{0},t) \cos^{2}(kx_{0}),
\end{equation}
where $T_{H}(x,x_{0},t)$ is given by Eq.(S33) in [36], leading to
\begin{equation}
	\Delta T_{\infty} = \Delta T_{\infty}^{0} \cos \Big[ 2 \pi (\bar{n}-1/2) \left( \dfrac{x-d_{z}/2}{d_{z}} \right)  \Big]; \quad 
	\Delta T_{\infty}^{0} \equiv \dfrac{d_{z}^{2} \alpha_{th}}{2 \pi^{2} \bar{n}^{2} k_{th}} I_{0}; \quad \bar{n} \equiv 2 n_{eff} \dfrac{d_{z}}{\lambda} + \dfrac{1}{2}. 
\end{equation}
Here, we assumed that the temperature increase vanishes at the ends of the metal substrate at $z=0,d_{z}$, $\lambda$ is subject to $n_{eff} d_{z}/ \lambda = (m-1/2)$ where $m$ is integer (the later stems from assuming that the intensity $I_{0} \cos^{2}(k(x-d_{z}/2))$ vanishes at the points $z=0,d_{z}$) and $n_{eff}$ is the effective refractive index of the corresponding mode. The corresponding TLD film deformation, analogous to Eq.(S35) in [36] is then given by
\begin{equation}
	\dfrac{\eta(x)}{h_{0}} = \alpha_{TC} \Delta T_{\infty}^{0} \cos \Big[\dfrac{n \pi (x-d_{z}/2)}{d_{z}} \Big]; \quad \alpha_{TC} \equiv \dfrac{3 \sigma_{T} d_{z}^{2}}{2 \sigma_{0} h_{0}^{2} \pi^{2}} \dfrac{(n-1/2)^{2}}{n^{4}}  ; \quad n \equiv 2 \bar{n} - 1.
\end{equation}

\end{document}